\documentclass[conference]{IEEEtran}

\usepackage[symbol]{footmisc}
\usepackage{mathtools}
\usepackage{bm}
\usepackage{esvect}
\usepackage{caption}
\usepackage{float}
\usepackage{amsfonts}
\usepackage{multirow}
\usepackage{url}
\usepackage{subfigure}
\usepackage{cuted}
\usepackage{hyperref}
\usepackage{ulem}
\usepackage{xcolor}

\newcommand{\exponent}{\mathsf{e}}
\newcommand{\significand}{\mathsf{d}}
\newcommand{\signbit}{\mathsf{b}}
\newcommand{\lap}{\mathsf{Lap}}

\newcommand{\RNum}[1]{\uppercase\expandafter{\romannumeral #1\relax}}

\newcommand{\gausstextbook}{\mathsf{Gauss\_polar}}
\newcommand{\gaussnumpy}{\mathsf{Gauss\_numpy}}
\newcommand{\gausspytorch}{\mathsf{Gauss\_pytorch}}
\newcommand{\gaussgo}{\mathsf{Gauss\_go}}

\newcommand{\difflabel}{\mathsf{DiffLabelCanary}}
\newcommand{\samelabel}{\mathsf{SimLabelCanary}}

\newcommand{\RandomDouble}{\mathsf{RandomFP}}
\DeclareTextFontCommand{\emph}{\textit}

\newcommand{\eps}{\epsilon}
\newcommand{\cd}{D}
\newcommand{\mech}{\mathcal{M}}

\newcommand{\RN}[1]{%
  \textup{\uppercase\expandafter{\romannumeral#1}}%
}

\begin{document}

\title{Are We There Yet? Timing and Floating-Point Attacks on Differential Privacy Systems}

\author{\IEEEauthorblockN{Jiankai Jin\IEEEauthorrefmark{1},
Eleanor McMurtry\IEEEauthorrefmark{2}\IEEEauthorrefmark{3},
Benjamin~I.~P. Rubinstein\IEEEauthorrefmark{1},
Olga Ohrimenko\IEEEauthorrefmark{1}}
\IEEEauthorblockA{\IEEEauthorrefmark{1}School of Computing and Information Systems, The University of Melbourne}
\IEEEauthorblockA{\IEEEauthorrefmark{2}Department of Computer Science, ETH Zurich}}

\newcommand{\noise}{\mathsf{NoiseDist}}
\newcommand{\IsFeasibleNormal}{\mathsf{IsFeasibleNormal}}
\newcommand{\IsFeasibleRandomDouble}{\mathsf{IsFeasibleRandomFP}}
\newcommand{\true}{\mathsf{true}}
\newcommand{\false}{\mathsf{false}}
\newcommand{\PolarFunction}{\mathsf{PolarMethod}}
\newcommand{\ZigguratFunction}{\mathsf{ZigguratMethod}}

\maketitle

\begin{abstract}
Differential privacy is a de facto privacy framework that has seen adoption
in practice via a number of mature software platforms. 
Implementation of differentially private (DP) mechanisms
has to be done carefully to ensure end-to-end security
guarantees.
In this paper we study two implementation flaws
in the noise generation commonly used
in DP systems. 
First we examine the Gaussian mechanism's
susceptibility to a floating-point representation attack.
The premise of this first vulnerability is similar to the one carried out by Mironov in 2011
against the Laplace mechanism. Our experiments show the attack's success against DP algorithms,
including deep learning models trained
using differentially-private stochastic gradient descent. 

In the second part of the paper we study discrete counterparts of the Laplace and Gaussian
mechanisms that were previously proposed to alleviate the shortcomings of floating-point representation
of real numbers. We show that such implementations unfortunately suffer from
another side channel: a novel timing attack. 
An observer that can measure the time to draw (discrete) Laplace or Gaussian
noise can predict the noise magnitude, which can then be used to recover sensitive 
attributes. This attack invalidates differential privacy guarantees
of systems implementing such mechanisms.

We demonstrate that several commonly used, state-of-the-art implementations 
of differential privacy are susceptible to these attacks.
We report success rates up to 92.56\% for floating point attacks on DP-SGD,
and up to 99.65\% for end-to-end timing attacks on private
sum protected with discrete Laplace.
Finally, we evaluate and suggest partial mitigations.
\end{abstract}

\section{Introduction}
\label{sec:intro}
\textit{Given the equation
\[z - y = 0.1234567890004\enspace,\]
can $y$ be equal to $0$, $2000$ or $20000$?
Though one may ask ``what is $z$?'', it is possible to answer this question without knowing $z$,
if one knows that the arithmetic was computed on a machine using the double-precision floating-point format.
While $z = 2000.1234567890004$ cannot be represented, 
 $2000.1234567890003$ and
$2000.1234567890006$ can be. Similarly for $z = 20000.1234567890004$. In fact without knowing $z$ at all, we can say definitively that $y$ must equal $0$ if it is known to be one of $0$, $2000$, or $20000$.}

\vspace{8pt}

\footnotetext[3]{Work done in part while at The University of Melbourne.}

Differential privacy (DP) is a de facto privacy framework that has received significant interest
from the research community and has been deployed by the U.S. Census Bureau, Apple, Google, Microsoft, and many others. 
Research on DP ranges from algorithms with different performance trade-offs,
to new models in different settings, and also to practical implementations~\cite{prochlo,dptflink,google-dp,opacus,ibm-gaussian}.
Robust implementations are crucial to provide end-to-end privacy 
that matches on-paper differential privacy guarantees.

Implementations of DP algorithms often raise concerns not considered in
theoretical analysis (which focuses on idealized settings).
Mironov~\cite{mironov} was first to discuss the implications of the fact that one cannot represent---and thus cannot sample from---all real numbers on a finite-precision computer.
Focusing on the Laplace mechanism, Mironov's attack proceeds by observing that certain floating-point values cannot be generated by a DP computation 
and hence a release could reveal the (private) noiseless value.
On the other hand, Haeberlen~\textit{et al.}~\cite{10.5555/2028067.2028100} and Andrysco~\textit{et al.}~\cite{DBLP:conf/sp/AndryscoKMJLS15} showed that DP algorithms may suffer from timing side-channels since such algorithms can take different time depending on sensitive values
in a dataset.

In this paper we extend Mironov's attack to other DP mechanisms and study its effects on real-world DP implementations. We then describe another timing side-channel that can arise
in DP implementations due to the timing of the noise samplers.

\paragraph{Floating-Point Representation of the Gaussian Distribution}
The \textit{Gaussian mechanism} (based on additive Gaussian noise) is another well-studied DP mechanism. 
It achieves what is often called \textit{approximate differential privacy}, 
meaning that the mechanism may fail completely to provide pure DP with some small and controllable probability $\delta$. 
However, the mechanism has advantages over the Laplace mechanism,
including lighter tails than the Laplace distribution and superior composition properties
when answering many queries with independent noise. Generalizations like R\'enyi differential privacy~\cite{renyi} 
can perform tight composition analysis. Recently \textit{truncated concentrated differential privacy}~\cite{bun2018composable} 
has emerged as a promising generalization that bounds the residual privacy loss from approximate DP, 
allows efficiently-computable optimal composition, and captures privacy amplification by subsampling as present in~\cite{abadi}.  
Because of these advantages, the Gaussian mechanism is often the tool of choice for applications such as deep learning~\cite{abadi} that 
involve carefully controlled privacy budgets over sequences of releases.

A question arises: are the same attacks as in~\cite{mironov}
possible against the Gaussian mechanism? Though several works~\cite{discrete,google_gauss} mention that it may be feasible,
to the knowledge of the authors no one has demonstrated this possibility nor shown how to carry out this attack in practice.
In this paper we study these two questions and demonstrate an attack confirming
that common implementations of Gaussian sampling are subject to floating-point attacks.

Unfortunately directly using the attack from~\cite{mironov} is not possible since Gaussian noise
is drawn using very different techniques to Laplace.
{The main challenge is due to some Gaussian samplers being based on two
random values and not one. Such methods produce two
independent Gaussian samples, and most implementations cache one of them to be used next time the mechanism is called. 
We develop an attack that uses both of these values. Due to the two equations with several unknowns
that the attacker needs to solve, there can be more than one pair of feasible values. As a result
our attack can yield false positives.} Nevertheless, we show that the attack is still feasible and
has a significant success rate.
Additionally, we show that a Gaussian sampler based on a different method
that generates only one sample is also susceptible to a floating-point attack, 
and the attack succeeds at a higher rate than for samplers based on two
values.

The most prominent recent use of the Gaussian mechanism is in training machine learning (ML)
algorithms using \textit{differentially-private stochastic gradient descent} (DP-SGD)~\cite{abadi}.
We show that we can mount the attack in this setting to determine if a batch contains a particular record or not,
violating DP guarantees of DP-SGD.
Moreover, ML model training naturally reveals sequential Gaussian samples to an adversary,
as it returns a noisy gradient for each parameter of the model.

\paragraph{Timing Attacks Against Discrete Distribution Samplers}
In the second part of the paper we study the primary method that has been proposed
to defend against floating-point attacks: discrete versions of the Laplace and Gaussian mechanisms~\cite{discrete,google_gauss}.
These approaches employ sampling algorithms that make no use of floating-point representations.
We observe that such mechanisms, though defending against
floating-point attacks, are susceptible to a timing side channel: an adversary who observes the time it takes to draw a sample
can determine the generated noise's magnitude. When used
within a DP mechanism, our attack reveals the noise contained in the result, and thus reveals the noiseless (private) value.

Our timing attack is possible due to the underlying technique
that these discrete samplers rely on: direct simulation of \textit{geometric sampling}, meaning values
are sampled until a coin toss results in a ``head''. 
The number of such coin tosses is tied to the magnitude of the noise returned; timing the sampler reveals this number and thus leaks the noise magnitude.
Though timing has been identified as a potential
side-channel in DP~\cite{10.5555/2028067.2028100,DBLP:conf/sp/AndryscoKMJLS15}, to our knowledge we are the first to show that noise distribution
samplers and not the mechanisms themselves give rise to secret-dependent
runtimes.

Our contributions are:
\begin{itemize}
\item We show that the Gaussian mechanism of differential privacy suffers from a side channel due to floating-point representation.
To this end, we devise attack methods to show how to exploit this vulnerability since the known floating-point attack against the Laplace mechanism cannot be used directly.
\item We use the above results to demonstrate empirically that Gaussian
samplers as implemented in \texttt{NumPy}, \texttt{PyTorch} and \texttt{Go} are vulnerable to our attack.
Focusing on the Opacus DP library implementation~\cite{opacus} by Facebook, we also show that DP-SGD is vulnerable to information leakage
under our attack.
\footnote{{Since notifying Facebook about the floating-point attack, Opacus library now 
has proposed a mitigation in \url{https://github.com/pytorch/opacus/pull/260}}}
\item We then observe that discrete methods developed to protect against floating-point attacks
for both the Laplace and Gaussian mechanisms suffer from timing side channels. We show that two libraries are vulnerable to these attacks:
a DP library by Google~\cite{google-dp}
and the implementation accompanying another work on discrete distributions in~\cite{discrete, discrete-imp}.
\item We discuss and evaluate mitigations against each attack.
\end{itemize}
\paragraph*{{Disclosure}}
{We have informed maintainers of the DP libraries mentioned above of the results of this paper. They have acknowledged
our report and notification of the disclosure dates.}
\section{Background}
\label{background}

In this paper we develop attacks on differential privacy
based on floating-point representation and timing channels.
In this section, we give background on how floating-point values are represented
on modern computers, differential privacy, and the Laplace and Gaussian
mechanisms for DP.

\subsection{Floating-Point Representation}
Floating-point values represent real values using three numbers:
a sign bit $\signbit$, an exponent $\exponent$,  and a significand $d_1d_2\ldots d_{\significand}$.
For example, 64-bit (double precision) floating-point numbers allocate 1 bit for $\signbit$, 11 bits for $\exponent$,
and 52 bits for the significand. Such a floating-point number is defined to be 
$(-1)^\signbit \times (1.d_1d_2\ldots d_{\significand})_2  \times 2^{\exponent-1023}$.

Crucially, the number of real values representable using floating-point values in the ranges $[a_1,b_1]$ and $[a_2, b_2]$, $a_i < b_i$, 
 are different even if $b_1-a_1 = b_2 - a_2$.
For example, there are approximately $2^{17}$ floating-point values in the range $[10, 10+2^{-32}]$,
and there are approximately $2^{14}$ floating-point values in $[100, 100+2^{-32}]$.
Thus, floating-point values are more densely distributed around 0.

\subsection{Differential Privacy}
Consider a collection of datasets $\mathcal{D}$ and an arbitrary space of output responses $\mathcal{R}$. We say that two datasets $\cd, \cd'$ are \textit{neighboring} if they differ on one record.
A randomized mechanism $\mech: \mathcal{D} \rightarrow \mathcal{R}$ is $(\eps,\delta)$-\textit{differentially private}~\cite{TCC06} for $\epsilon>0$ and $\delta\in[0,1)$ if given any two neighboring datasets
$\cd, \cd' \in \mathcal{D}$ and any subset of outputs $R \subseteq \mathcal{R}$
it holds that $\Pr[\mech(\cd) \in \mathcal{R}] \le e^\eps \Pr[\mech(\cd') \in \mathcal{R}] + \delta$.
That is, the probability of observing the same output~$y$ from $\mech(\cd)$ and $\mech(\cd')$ is bounded.
DP therefore guarantees that given an output~$y$,
an attacker cannot determine which of $\cd$ or $\cd'$ was used for the input.
If there is some output that is possible with $\cd$ but not $\cd'$, this inequality cannot
hold for non-trivial $\delta$, so the mechanism cannot be DP. This is the key fact our attacks exploit.

In this paper, we consider DP mechanisms $\mech$ that provide protection
by computing the intended function $f$ on the data $\cd$
and randomizing its output.  That is, $\mech(\cd) = f(\cd) + s$ 
where $s \leftarrow \noise$ is noise drawn either from Laplace or Gaussian 
distributions with appropriate parameters (as discussed below). 
For example, $f$ may compute the sum of the set of employee incomes $\cd$, 
and we may wish to keep the exact values of the incomes private.
Our attacks are based on observations about $\noise$ that can be used
to learn the noise sampled from it.

\subsection{Laplace Mechanism}
\label{sec:laplace}
The Laplace mechanism provides $\epsilon$-DP by additive noise drawn from the 
Laplace distribution $\lap(\lambda)$ (i.e., $\noise$ is $\lap(\lambda)$). 
In the scalar-valued case $\mech(\cd) = f(\cd) + \lap(\lambda)$.
Here we use a scale parameter $\lambda = \Delta/\eps$ where $\Delta$ is $f$'s sensitivity, 
meaning any neighboring data sets $D$ and $D'$ satisfy $|f(\cd) - f(\cd')| \le \Delta$.

\subsection{Gaussian Mechanism}
The Gaussian mechanism~\cite{privacybook} provides $(\eps, \delta)$-DP to the 
outputs of a target 
function $f: \mathcal{D} \rightarrow \mathcal{R}$, where $\mathcal{R}=\mathbb{R}^d$. 
The mechanism is popular due to its favorable noise tails 
compared to alternative mechanisms, and its 
composition properties when the mechanism is used to answer
many queries on data, as is common
when training a machine learning model or answering repeated queries
on a database~\cite{DBLP:journals/corr/DworkR16,renyi}.

Let $\Delta_f$ be the $L_2$-\textit{sensitivity} of $f$,
that is, the maximum distance $\left\lVert  f(\cd) - f(\cd')\right\rVert_2$
between any neighboring datasets $\cd$ and $\cd'$. Then the Gaussian mechanism 
$\mech(\cd)$ adds noise $\noise$ to $f(\cd)$. We write 
$\mathcal{N}(x, \sigma^2\Delta_f^2\mathbb{I})$ to mean the multivariate Gaussian with mean 
given by the target $x$ and covariance given by the identity matrix scaled by 
$\sigma^2\Delta_f^2$. We then have $\noise = \mathcal{N}(0, \sigma^2\Delta_f^2\mathbb{I})$, 
and the output distribution is $\mathcal{N}(f(D), \sigma^2\Delta_f^2\mathbb{I})$. 
The resulting mechanism is $(\eps, \delta)$-DP
if $\sigma =  \sqrt{2\log (1.25/\delta)}/\eps$ for arbitrary $\eps>0$ and $\delta\in(0,1)$.

In addition to applications computing a one-off release of a function output, 
the Gaussian mechanism is commonly used repeatedly in training
machine learning models using mini-batch stochastic gradient descent (SGD).
This composite mechanism is called DP-SGD~\cite{Bassily:2014:PER:2706700.2707412,6736861}.
When used to replace non-private mini-batch SGD, it
produces a machine learning model with
differential privacy guarantees on sensitive training data.
This mechanism has been applied
in Bayesian inference~\cite{pmlr-v37-wangg15},
to train deep
learning models~\cite{abadi, DBLP:conf/iclr/McMahanRT018, dpmp_dl},
and also in logistic regression models~\cite{6736861}.
At a high level, a record-level DP-SGD mechanism aims to protect presence of a record
in a batch and, hence, in the dataset.
DP-SGD has been also used in the Federated Learning setting~\cite{DBLP:conf/iclr/McMahanRT018}
where
each client computes
a gradient on their local data batch, adds noise and sends the result to a central server.

\section{Threat Model}
\label{sec:threat}
We consider an adversary that obtains an output of an implementation
of a differentially-private mechanism, for example a DP-protected average income of people in a database of personal records,
or DP-protected gradients used to train a machine learning model.
DP aims to defend {information about} presence (or absence) of a certain record
by providing plausible deniability. 
We show that the attacker can use artefacts of implementations of noise samplers in DP mechanisms
to undermine their guarantees.
In particular we consider two attacks based on separate artefacts --- one on floating-point representation
and one on timing.

We envision three scenarios where our attacks can be carried out:
\begin{enumerate}
\item[S:] a member of the public observes \textit{statistics} computed on sensitive data and protected with DP noise (e.g., those released by the US Census Bureau~\cite{10.1145/3219819.3226070});
\item[DB:] an analyst interactively queries a differentially private \textit{database}~\cite{Johnson:2018:TPD:3187009.3177733,McSherry:2009:PIQ:1559845.1559850,10.5555/2028067.2028100} which allows them to ask several queries of
datasets and which transparently adds noise to preserve
privacy of the dataset from the analyst;
\item[FL:] a central server who is coordinating \textit{federated learning} by collecting
gradients from clients~\cite{DBLP:conf/iclr/McMahanRT018}.
Here, a client adds noise to a gradient computed on its data to protect its data from the central server.
\end{enumerate}

For all three scenarios above, our threat model builds
on the threat model of DP where (1) the attacker observes a DP-protected output,
(2) the attacker may know all the other records in the dataset except
for the one record it is trying to guess, and (3)
knows how the mechanism is implemented
\footnote{Many DP implementations are open-source including~\cite{dptflink,google-dp,opacus,ibm-gaussian}.}
, but does not know the randomness used by it.

We describe additional adversarial capabilities required for each scenario and attack below.

\paragraph{{Floating-Point Attack (Section~\ref{sec:gaussimplem})}}
\label{sec:sample}
For one of our two floating-point attacks, {in addition to observing a single DP output that the adversary wishes to attack},
we assume the adversary has access to an output of a consecutive execution of a DP mechanism or its
noise sampling.
This is achievable in practice in all scenarios above due to:
\begin{enumerate}
\item \emph{multiple queries:} In scenario S multiple statistics are released,
in scenario DB a (malicious) analyst could query a DP protected database several times.
\item \emph{$d$-dimensional query:} in all three scenarios, an output being protected 
can correspond to an output of a $d$-dimensional function where independent noise is added 
to each component such as a histogram~\cite{privacybook} or a gradient computed for multiple 
parameters of ML model~\cite{abadi}. Moreover, gradients are assumed to be revealed as part 
of the privacy analysis in the central setting as well as in the FL setting.
\end{enumerate}

\paragraph{Timing Attack (Section~\ref{sec:timingtheory})}
In contrast to the floating-point attack above, here the threat model assumes that 
the attacker observes a DP output \textit{and} can measure the time
it takes for the DP mechanism to compute {it}.
The attacker must also be able to measure multiple runs of an algorithm in order to obtain a baseline
of running times.
However during the attack itself, the attacker only needs to make one observation to make a reasonable guess.

The attack can be deployed in the three scenarios above if
an attacker has black-box access to the machine running
DP code, similarly to the threat model of other timing side-channels used 
against DP mechanisms that are \textit{not} based on noise 
samplers~\cite{DBLP:conf/sp/AndryscoKMJLS15,10.5555/2028067.2028100} 
(see Section~\ref{sec:related} for more details). For example, the 
attacker may share a machine based in the 
cloud~\cite{prochlo,DBLP:journals/corr/abs-1807-00736,Ristenpart:2009:HYG:1653662.1653687} 
or is a cloud provider itself.
For the DB scenario specifically, a malicious analyst querying the mechanism 
hosted locally can readily measure the time it takes for the query to return.
For the FL scenario (and remotely hosted databases in the DB scenario) the uncertainty in measuring
precise time due to network communication can be reduced with recent attacks 
exploiting concurrent requests~\cite{10.5555/3489212.3489324}.
\section{Floating-Point Attack on Normal Distribution Implementations}
\label{sec:fpattack}

We describe the floating-point attack that aims to determine
whether a given floating-point value
could have been generated by an instance of a Gaussian
distribution or not. If not, this eliminates the possibility that a DP mechanism
could have used this noise, hence undermining its privacy guarantees.
We begin with a description of a generic floating-point attack against DP
and then describe two common implementations of Gaussian samplers---polar and Ziggurat---and how they can be attacked.

\subsection{Floating-Point Attack {on DP}}
\label{sec:fpattackdp}

The DP threat model assumes that the adversary
knows neighboring datasets $\cd$, $\cd'$ and function~$f$.
Given an output~$y$ of a DP mechanism,
where either 
$y = f(\cd) + s$ or
$y = f(\cd') +s'$ and $s,s' \leftarrow  \noise$,
the attacker's goal is to determine if $\cd$ or $\cd'$ was used 
in the computation of~$y$.

Mironov~\cite{mironov} showed that due to an artefact in the implementation
of $\noise$ for Laplace, some values of $s$ are impossible.
Hence given $y$, if the adversary knows that $s$ is impossible
then it must be the case that $\cd'$ was used to compute $y$ (and similarly for $s'$).
This directly breaks the guarantee of DP
which states that there is a non-zero probability
for each of the inputs producing the observed output.
We will show that mechanisms that use Gaussian noise for $\noise$
--- whose implementation is more complicated than Laplace --- 
are also susceptible to implementation artefacts.

In the rest of this section we develop
a function $\IsFeasibleNormal(s)$ which returns $\true$ if a given noise value~$s$ could
have been drawn from implementations of Gaussian distributions
and $\false$ otherwise.
The attacker then runs $\IsFeasibleNormal(s)$ and $\IsFeasibleNormal(s')$.
If only one of them returns $\true$, the attacker determines that the corresponding
dataset was used in the computation. Otherwise, it makes a random guess.

\subsection{Warmup: Feasible Random Floating Points}
\label{sec:randomfp}
We describe how random double-precision floating-point values (``doubles'') are sampled
on modern computers using a function $\RandomDouble$
and show that given a double $x$, one can determine if it was generated using $\RandomDouble$
or not.
This will serve as a warm-up for our attack against the Gaussian distribution over doubles.

Random real values in the range {$(0,1)$} can be drawn by choosing a random integer {$ u$ from $[1,R)$} and
then dividing it by the resolution $R=2^{p}$, where the value of $p$ varies by system.
We abstract this process using a function $\RandomDouble$ that
chooses an integer $u$ at random from $[1,R)$ and returns $u/R$.
Given a double $x$ one can determine if it
could have been produced by $\RandomDouble$ by checking if
$ x \stackrel{?}{=} \bar{u} /R$
for some integer $\bar{u} \in [1, R)$.
If the equality holds then $x$ could have been
produced from a random integer.
Since rounding errors are introduced during multiplication
and division, we will later also perform the above check for neighboring values of $x$.

\subsection{Polar Method: Implementation and Attack}

In this section we describe the \textit{polar method} and the floating-point attack against it.
\label{sec:gaussimplem}

\subsubsection{Method}

\label{sec:polar}
The \textit{Marsaglia polar method}~\cite{marsaglia_polar} is a computational method
that generates samples of the standard normal distribution from uniformly distributed random values.
$\PolarFunction$ operates as follows:
\begin{enumerate}
  \item[\bf{P1}] Choose independent uniform random values $x'_1$ and $x'_2$ from $(0, 1)$
  {using $\RandomDouble$}.
  \item[\bf{P2}] Set $x_1 \leftarrow 2x'_1-1$ and $x_2 \leftarrow 2x'_2-1$. (Note that both fall in the interval $(-1, 1)$.)
  \item[\bf{P3}] Set $r\leftarrow x_1^2+x_2^2$.
  \item[\bf{P4}] {Repeat from Step {\bf P1} until $r \leq 1$ and $r \neq 0$.}
  \item[\bf{P5}]  Set $s_1 \gets x_1\sqrt{\frac{-2\log{r}}{r}}$ and $s_2 \gets x_2\sqrt{\frac{-2\log{r}}{r}}$.
   \item[\bf{P6}] Return $s_1$.
\end{enumerate}
The procedure generates two independent samples from a normal distribution:
$s_1$ and $s_2$. 
{T}he second value{, $s_2$,} {is cached} and return{ed} on the next invocation.
If the cache is empty, the sampling method is invoked again.
The method can generate samples from $\mathcal{N}(0,\sigma^2)$
by returning $\sigma s_1$ and $\sigma s_2$ instead.

The polar method is used by both the GNU C++ Library with \texttt{std::normal\_distribution}
and the Java class \texttt{java.util.Random} (in the \texttt{nextGaussian} method).
A related technique called \textit{Box-Muller method} 
described in the Appendix~\ref{sec:bm} that relies on computing $\sin$
and $\cos$ is used in PyTorch and {was implemented
in the older versions of Diffprivlib~\cite{ibm-gaussian}}.

\subsubsection{Floating-Point Attack}
\label{sec:fpattacknormal}

The attacker's goal is to devise a function $\IsFeasibleNormal(s)$
that determines if value $s$ could
have been generated by $\PolarFunction$ --- a computational method for drawing normal
noise. Here, we assume that the attacker knows sample
$s_2$ and is trying to guess if
$s \stackrel{?}{=}s_1$.
As discussed in~Section~\ref{sec:threat}, an attacker can learn $s_2$ either through
multiple queries or multi-dimensional queries.
We note that the attack against the Laplace method by Mironov~\cite{mironov} 
cannot be applied since $\PolarFunction$ (and the Box-Muller method)
(1) relies on different mathematical formulae and (2)
uses two random values and draw two samples from their target distribution.
To this end, we devise an FP attack
specifically for normal distribution implementations.
We describe the attack for the polar method below.
The attack for the Box-Muller method proceeds similarly, using trigonometric functions instead.

Before proceeding with the attack, we observe that $r$ in $\PolarFunction$ can
be expressed using $s_1$ and $s_2$ by simple arithmetic rearrangements
based on steps {\bf P3} and {\bf P5}.
\begin{equation*}
  \begin{aligned}
    r &=  x_1^2 + x_2^2  
  = \left( s_1 \sqrt{\frac{r}{-2\log r }} \right)^2 +\left(s_2 \sqrt{\frac{r}{-2\log r}}\right)^2  \\ &=
     \frac{s_1^2 r}{{{-2\log(r)}}} +\frac{s_2^2 r}{{{-2\log(r)}}} 
  \end{aligned}
\end{equation*}

Rearranging this equation further, we obtain

\begin{equation}
  \begin{aligned}
    -2 \log r &= 
     {s_1^2} +{s_2^2} \\
     r &=  \exp\left(- \frac{{s_1^2} +{s_2^2} }{2} \right) \label{eq:rcomp}
  \end{aligned}
\end{equation}

The intuition behind our attack is similar to that of ``attacking'' $\RandomDouble$
in Section~\ref{sec:randomfp}: we find an
expression that must be an integer if $\PolarFunction$ was used.
We observe that $r \times R^2$ must be an integer:
since $x'_1$ and $x'_2$ are produced using $\RandomDouble$ (step~{\bf P1})
 there must be integers $u_1$ and $u_2$ such that
$x'_1 = u_1 / R$ and $x'_2 = u_2 /R$.
Rearranging and substituting these $x'_1$ and $x'_2$ further in steps {\bf P2} and  {\bf P3}
we obtain
\begin{equation*} 
r = (2 u_1 / R - 1)^2 + (2 u_2 /R - 1)^2 
\end{equation*}
Since $u_1$, $u_2$, $R$ are integers,  value $r \times R^2$
must be an integer.

$\IsFeasibleNormal(s)$ proceeds as follows.
The attacker computes value of $r \times R^2$
using Equation~\ref{eq:rcomp} 
with values $s$ (instead of $s_1$) and~$s_2$ as follows:
\begin{equation} \exp\left(- \frac{{s^2} +{s_2^2} }{2} \right) \times R^2
\label{eq:fpattack}
\end{equation}
It then checks 
if the value in Equation~\ref{eq:fpattack} is an integer. If it is, 
then $\IsFeasibleNormal(s)$ returns $\true$ since
$s$ and $s_2$ could be produced using $\PolarFunction$.

As floating-point arithmetic cannot be done with infinite precision
on finite machines, $s$ and $s_2$
might be inaccurate. To this end, we perform a heuristic search in each direction
of $s$ and $s_2$ where we
try several neighboring values of $s$ and~$s_2$.
{In our experiments, we choose to search 50 values in each direction.}
This search heuristic is prone to errors and may result
in false positives and false negatives, due to 
(1) more than one pair of values resulting in $s_1$ and $s_2$ and
(2) our search not being exhaustive in exploring values with a range of 100 values.
Nevertheless in the next section we show that the attack is still successful
for a variety of applications of Gaussian noise in DP. 

\subsection{Ziggurat Method: Implementation and Attack}
\label{sec:zigimplem}

The Ziggurat method~\cite{marsaglia_tsang_2000} is another method for generating
samples from the normal distribution.
Compared to the Box-Muller and polar methods, it generates one sample on each invocation.
It is a rejection sampling method that randomly generates a point in a distribution slightly 
larger than the Gaussian distribution. It then 
tests whether the generated point is inside the 
Gaussian distribution.

\subsubsection{Method}

$\ZigguratFunction$ relies on three precomputed tables
$w[n]$, $f[n]$ and $k[n]$
that are directly stored in the source code. 
The Ziggurat implementation in Go uses $n=128$ and proceeds as follows.
\begin{enumerate}
  \item[\bf{Z1}] Generate a random 32-bit integer $j$ and let $i$ be the index provided by
  the rightmost 7 bits of $j$.
  \item[\bf{Z2}] Set $s\leftarrow jw[i]$. If $j<k[i]$, return $s$.
  \item[\bf{Z3}] If $i=0$, run a fallback algorithm for generating a sample from the tails of the distribution.
  \item[\bf{Z4}] Use $\RandomDouble$ to generate 
    independent uniform random value $U$ from $(0, 1)$.
    Return $s$ if: $$U \left(f[i-1]-f[i]\right) < f(s)-f[i]$$
  \item[\bf{Z5}] Repeat from step \textbf{Z1}.
\end{enumerate}
We refer interested readers to~\cite{ziggurat_tail} for how to sample from the tail of a
normal distribution and to~\cite{marsaglia_tsang_2000} for
how the tables $w[n], f[n], k[n]$ are generated.
The method can generate samples from $\mathcal{N}(0,\sigma^2)$
by returning $\sigma  s$.

The Ziggurat method is used by the
Go package \texttt{math/rand} with \texttt{NormFloat64}
and new random sampling methods of 
\texttt{NumPy}\footnote{\url{https://numpy.org/doc/stable/reference/random/index.html}}.

\subsubsection{Attack}
\label{sec:gaussimplemZ}
This time, the attacker aims to devise a function $\IsFeasibleNormal(s)$
that determines if value $s$ could
have been generated by $\ZigguratFunction$.
We describe our attack steps below since attacks in Section~\ref{sec:gaussimplem} 
and~\cite{mironov} are not applicable due to pre-computed tables used in Ziggurat.

We observe that the returned value is $jw[i]$ in steps {\bf Z2} and~{\bf Z4}, 
except when the value is sampled from the tail of Gaussian distribution in step {\bf Z3} (which happens infrequently).
Furthermore, $j$ is an integer
and all values of $w[n]$ are available to the attacker, because 
all precomputed tables
are stored directly in the source code. 
Based on these observations, 
for a noise $s$, our attack for $\ZigguratFunction$ proceeds as:
\begin{enumerate}
    \item For each $w[i]$, $i\in[1,n]$ calculate $\mathsf{w}[i]={s}/(\sigma w[i])$.
    \item For each $\mathsf{w}[i]$, check if it is an integer.
    \item If any $\mathsf{w}[i]$ is an integer, then we take $s$ as a feasible 
    floating-point value, $\IsFeasibleNormal(s)$ returns $\true$.
\end{enumerate}
Note that the method does not attack values sampled from the tail.
However, we observed that it happens less than $0.1\%$ of the time when $n=128$,
hence, the attack works for the majority of the cases.

\section{Experiments: FP Attack on Normal Distribution}
We perform four sets of experiments to evaluate leakage of Gaussian samplers
based on the attack described in the previous section.
First, we enumerate the number of possible values
in a range of floating points that can result in an attack and observe that it is not negligible.
We then show the effectiveness of our attack on private count and DP-SGD, 
a popular method designed for training machine learning (ML) models 
under differential privacy~\cite{abadi}.

\paragraph*{Gaussian Implementations}

In our experiments, we use four implementations of Gaussian distribution samplers.
\begin{itemize}
  \item $\gausstextbook$: our implementation of polar method as described in Section~\ref{sec:polar}.
  \item $\gaussnumpy$: \texttt{NumPy} implementation of polar Gaussian sampling.
  \item $\gausspytorch$: \texttt{PyTorch} implementation of Box-Muller Gaussian sampling.
  \item $\gaussgo$: \texttt{Go} implementation of Ziggurat Gaussian sampling.
\end{itemize}
Our attacks on DP-SGD were tested using the privacy engine implementation of 
the Opacus library~\cite{opacus}\footnotemark[1].

\subsection{Distribution of ``Attackable'' Values}
\label{sec:fpattackimp}
In this section we aim to understand how many floating-point values
could not have been generated from a Gaussian sampling implementation.
That is, for how many values~$s$,~$\IsFeasibleNormal(s)$ would return $\mathsf{false}$.
We call such values ``attackable'' since if an adversary were to observe them,
they would know that the value must have been produced in a specific manner (e.g.,
by adding noise to $f(D')$ as opposed to $f(D)$).
We use the $\gausstextbook$ implementation to perform this experiment in a controlled environment
and assume that the adversary is given $y_1$ and $y_2$
where $y_1 = f(D) + s_1$ and $y_2 = f(D) + s_2$ or $y_1 = f(D') + s_1'$ and $y_2 = f(D') + s_2'$.
(Compared to the attack in Section~\ref{sec:fpattack} the attacker is not given~$s_2$ directly,
however the attack can proceed similarly.)

\begin{figure}[t]
  \includegraphics[width=0.5\textwidth]{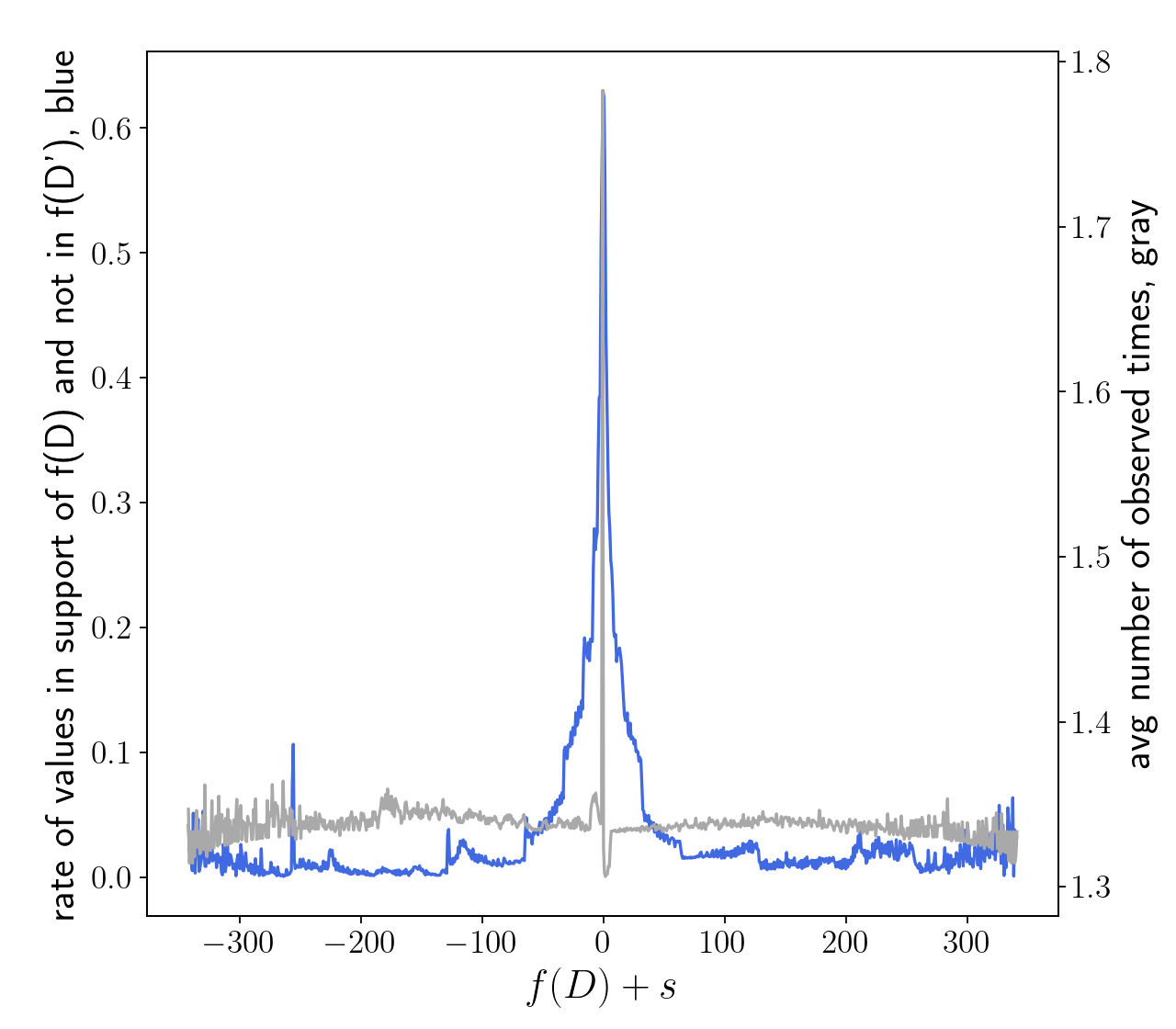}
  \centering
  \caption{
    Distribution and number of values amenable to a floating-point attack against a Gaussian
    implementation $\gausstextbook$ using $R=2^{10}$ and $\sigma=114$. Here the horizontal axis, $f(D) +s$, shows both values $f(D) + s_1$ and $f(D) + s_2$.
    For presentation purposes, rate of attackable floating-point values is averaged over $0.8$-wide intervals.
    Blue line indicates the rate of floating-point values that are in the support
    of $f(D) = 0$ and not in $f(D') = 1$.
    Gray line measures average number of times support for $f(D)$ was observed.
  }
  \label{fig:gaussian_values_attack}
\end{figure}

For each pair of $y_1 = f(D) + s_1$ and $y_2 = f(D) + s_2$, 
if the attack successfully concludes that they are in support 
of $f(D)$, and not $f(D')$, we count them as attackable values.
We set $f(D) =0$, $f(D') = 1$, $\sigma = 114$, $R=2^{10}$.
The blue line 
of Figure~\ref{fig:gaussian_values_attack} shows the distribution of the
rate of attackable values where we plot~$y_1$ and $y_2$.
In order to fit all measured floating-points into the resolution of the graph,
we aggregate the results over small intervals of width $0.8$.

Mironov generated a similar graph for the Laplace distribution in~\cite{mironov}.
Compared to Gaussian, the
Laplace distribution has more attackable values since it
uses only one random value, hence the attack does not suffer from false negatives.

In order to understand our results further we also plot the average number of times
each $y=f(D)+s$ is observed, using the gray line of Figure~\ref{fig:gaussian_values_attack}. 
This explains some of the spikes in the blue line: more frequently observed values
indicate more attackable values.
Percentage of floating-points that are attackable are higher closer to $0$
(recall that the graph shows an average over FP intervals).
Overall, the graph suggests that attackable values do exist
and as we show in the rest of this section
provide an avenue for an attack against DP mechanisms.

\subsection{DP Gaussian Mechanism}
\label{float}

\subsubsection{{Private count}}
\label{sec:privatecountFP}

\begin{figure}[t]
  \includegraphics[width=0.5\textwidth]{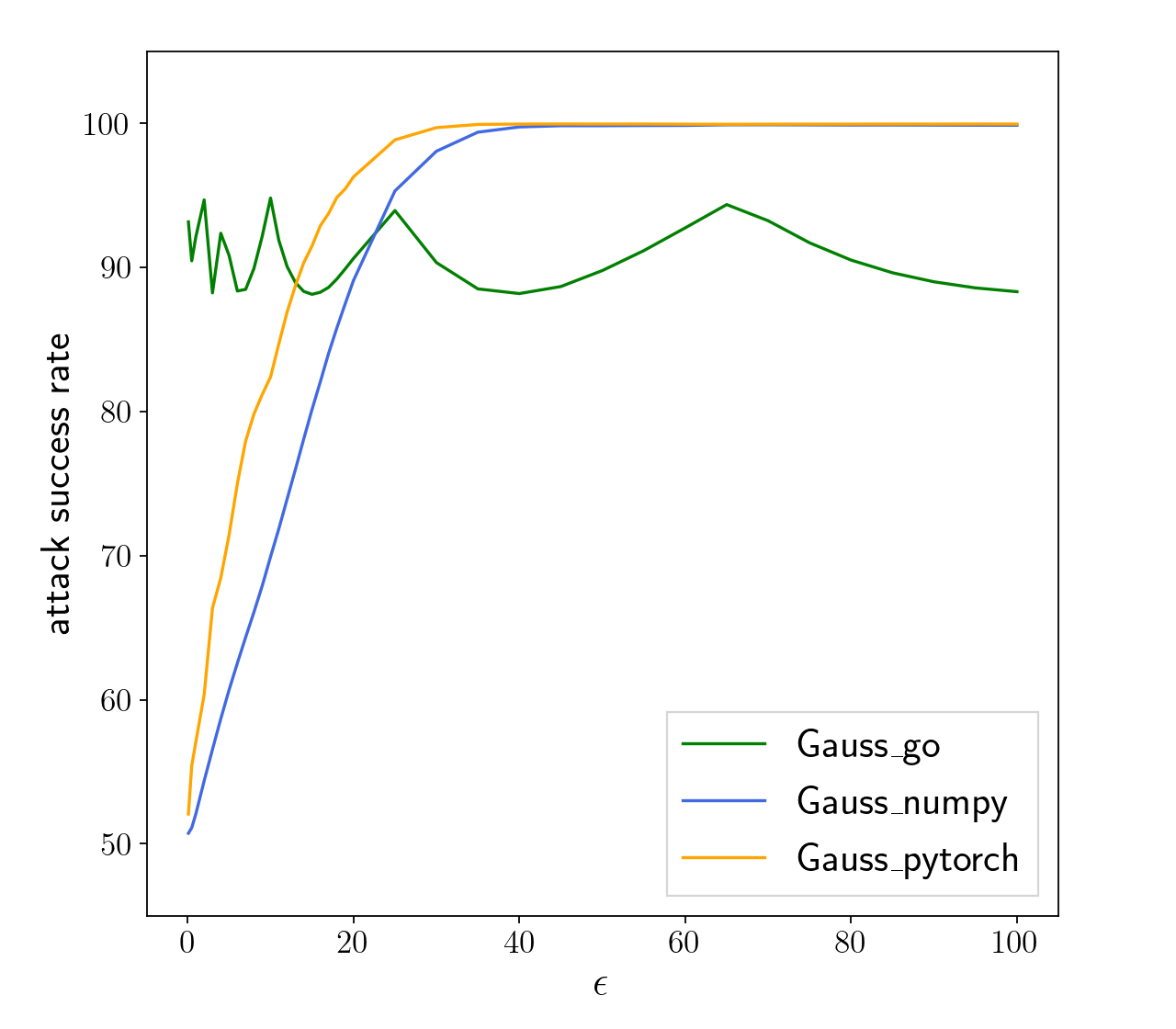}
  \centering
  \caption{
    FP attack success rate on private count
    where the count is protected with one of the three Gaussian samplers
    across $\epsilon \in (0,100]$ with fixed $\delta=10^{-5}$
    and function sensitivity $\Delta=1$.
    Baseline (random) attack success is~50\%.} 
  \label{fig:eps-succ}
\end{figure}

In this section, we explore the effectiveness of our attack against the Gaussian mechanism used
to protect a private count (i.e., corresponding
to S or DB scenarios in Section~\ref{sec:threat}).
We use the German Credit Dataset~\cite{german-credits}
and the query: count the number of records with number of credits 
greater than 16K (resulting in one record since the two
highest values among dataset's 1000 records are 15945 and 18424).

Given the output of the Gaussian mechanism, $y$, the adversary's
goal is to determine whether
noiseless output came from $q = f(D)$ or $q' = f(D')$ where $f$ is the count query defined above.
The private count of values in a dataset is computed as:
$ y = q + s $ or $ y = q' + s $,
where $q$ and $q'$ are the outputs of $f$ on dataset $D$ and $D'$, respectively, 
and~$s$ is a noise sampled with $\gaussnumpy$, $\gausspytorch$, or $\gaussgo$.
For $\gaussnumpy$ and $\gausspytorch$, the adversary also knows the second
value sampled from the distribution (i.e., $s_2$ in Section~\ref{sec:gaussimplem}).
The neighboring datasets $D$ and $D'$ 
that our attack tries to distinguish
differ on a single record whose credits is 0 in $D$ and 18424 in $D'$,
so the count of records satisfying the above query is $q =0$ for $D$
and $q'=1$ for $D'$.

We set the sensitivity $\Delta=1$ since one record can change the output of $f$ only by 1.
We vary $\epsilon$ in the range $(0,100]$ with fixed $\delta=10^{-5}$.
For each tuple of $\epsilon$, $\delta$ and $\Delta$, we use 
the analytic Gaussian Mechanism~\cite{analytic_gaussian} to calculate the required noise scale $\sigma$ for the 
experiment. 
Here, small $\epsilon$ and $\delta$ model a common set of parameters used in DP~\cite{privbook}.

In Figure~\ref{fig:eps-succ} we plot our success rate from 1 million trials from the following attack.
Recall that the attacker cannot always make a guess due to the limitations 
listed in~Section~\ref{sec:fpattackimp}. To this end,
if the attack can find support for either only $q$ or only $q'$, then 
the attack outputs the corresponding guess.
If the result is in support of both or neither of $q$ and $q'$, 
then the attacker is unsure and resorts to the baseline attack, choosing $q$ or $q'$ at random.

We observe that the attack is more successful for the $\gaussgo$ method for
values of $\epsilon$ less than 10. This is also the range of values 
for $\epsilon$ used in the literature on DP~\cite{hsu2014differential}. 
We note a slight cyclic behavior and observe that
peaks occur when the corresponding noise scale $\sigma$ (w.r.t.~$\epsilon$) is a power of 2
(see Figure~\ref{fig:zig_pattern} in the Appendix). 
We hypothesize that the success rate at those points is higher since multiplication 
and division by powers of 2 can be done via bit shifting that decreases the impact of rounding. 
Hence, extraction of $s$ is not affected by rounding errors that would otherwise 
be introduced during the multiplication of $s$ by $\sigma$ in the 
$\ZigguratFunction$ and step 1 of the corresponding attack in Section~\ref{sec:zigimplem}.

Among two sampler methods, attack is more efficient against $\gausspytorch$ than $\gaussnumpy$.
We also observe that the attack becomes stronger as $\epsilon$ increases.
This is potentially due to the higher distribution of attackable
values as indicated in Figure~\ref{fig:gaussian_values_attack}.
The attack is always more successful than the baseline random attack.

In Appendix~\ref{app:recall} we also investigate the rate at which the attacker 
can make a guess (attack rate) and how many of these guesses are correct (attack accuracy)
across a range of $\epsilon$ and sensitivity values.
$\gaussnumpy$ and $\gausspytorch$ have attack rates ranging between
$1.7\%$ and $92.8\%$ with attack accuracy of at least $89\%$.
In comparison, the attack rates and accuracy against $\gaussgo$ are at least~$76\%$
and~$99\%$, respectively.

\subsection{DP-SGD in Federated Learning scenario}
\label{sec:dpsgdattack}
The Gaussian mechanism is used extensively in training ML
models via differentially-private stochastic gradient descent~\cite{abadi,dpmp_dl,pmlr-v37-wangg15}. 
In this section, we describe successful attacks on the differentially-private training of ML models.
Here, we consider the Federated Learning scenario  from Section~\ref{sec:threat}
where an attacker (central server) observes
a DP-protected gradient computed on a batch of client's data
and is trying to determine if the batch includes a particular record or not.

The main observations we make in this section are:
\begin{itemize}
\item An adversary can determine if a batch contains a record with a different label from other records in the batch, i.e.,
a record that comes from the same distribution as training dataset but different to those in the batch.
\item The success rate of the attack increases as $\eps$ decreases as there are fewer floating-point values to represent large noise values.
\end{itemize}

\subsubsection{Setup}
We use Opacus library~\cite{opacus} for training machine learning models
with DP-SGD.
Our experiments are based on the Opacus example on MNIST data.
This dataset~\cite{mnist} contains 70K images,
split in 60K and 10K sets for training and testing, respectively. Each image is in $28\times 28$
gray-scale representing a handwritten digit ranging from 0 to 9 where the written digit is the label of the image.
The ML task is: given an unlabelled image, predict its label.

We use the same parameters for preprocessing and neural 
network training as used in the library~\cite{DBLP:journals/corr/abs-1911-11607,opacus}, following~\cite{abadi}. 
These training parameters are: learning rate $0.1$, epoch number~$8$, batch size $S=64$, 
number of batches in each epoch 100, and DP-SGD clipping norm~$L=1$.
We vary $\epsilon$ to understand how different parameters affect
the attack success rate while keeping a fixed  $\delta=10^{-5}$ 
(as in~\cite{DBLP:journals/corr/abs-1911-11607,abadi}).
Depending on the $\epsilon$, the noise is drawn from a normal distribution
with $\mu=0$ and 
$\sigma\in[1,250]$.
As a result $\epsilon$ ranges from 0.1 to 0.68.
Since the model has $d = 26,010$ parameters, each batch is used in $d$ gradient computations
to update the corresponding parameters. Hence, the attacker obtains $d$ floating-point values
protected with DP-SGD.
Details of DP-SGD computation are provided in Appendix~\ref{app:gradcomp}.

We consider a setting where an adversary observes a gradient computed on FL client's
batch of labelled MNIST images. The batch could correspond either to records $B$ or a neighboring batch 
$B'$ produced by replacing a randomly chosen 
record in batch $B$ with a \textit{canary record} (defined below).
The attacker's goal is, given $B$, $B'$ and a noisy gradient protected with Gaussian noise,
to determine if the gradient was computed on $B$ or $B'$.

We use different types of batches and canary records 
to test the effectiveness of our attack:
\begin{itemize}
  \item $\samelabel$:  batch $B$ is composed of shuffled training data.
  The canary record of its neighboring batch $B'$ is a record drawn from the test dataset.
  \item $\difflabel$: the batch is composed of records with the same labels in $[1,9]$, and
  the canary record of its neighboring batch is a record with label $0$.
\end{itemize}

Though $\difflabel$ is handcrafted, it represents an example of a batch that should be protected
by DP guarantees since all records are drawn from the same distribution.

\begin{figure}
  \includegraphics[width=8.7cm]{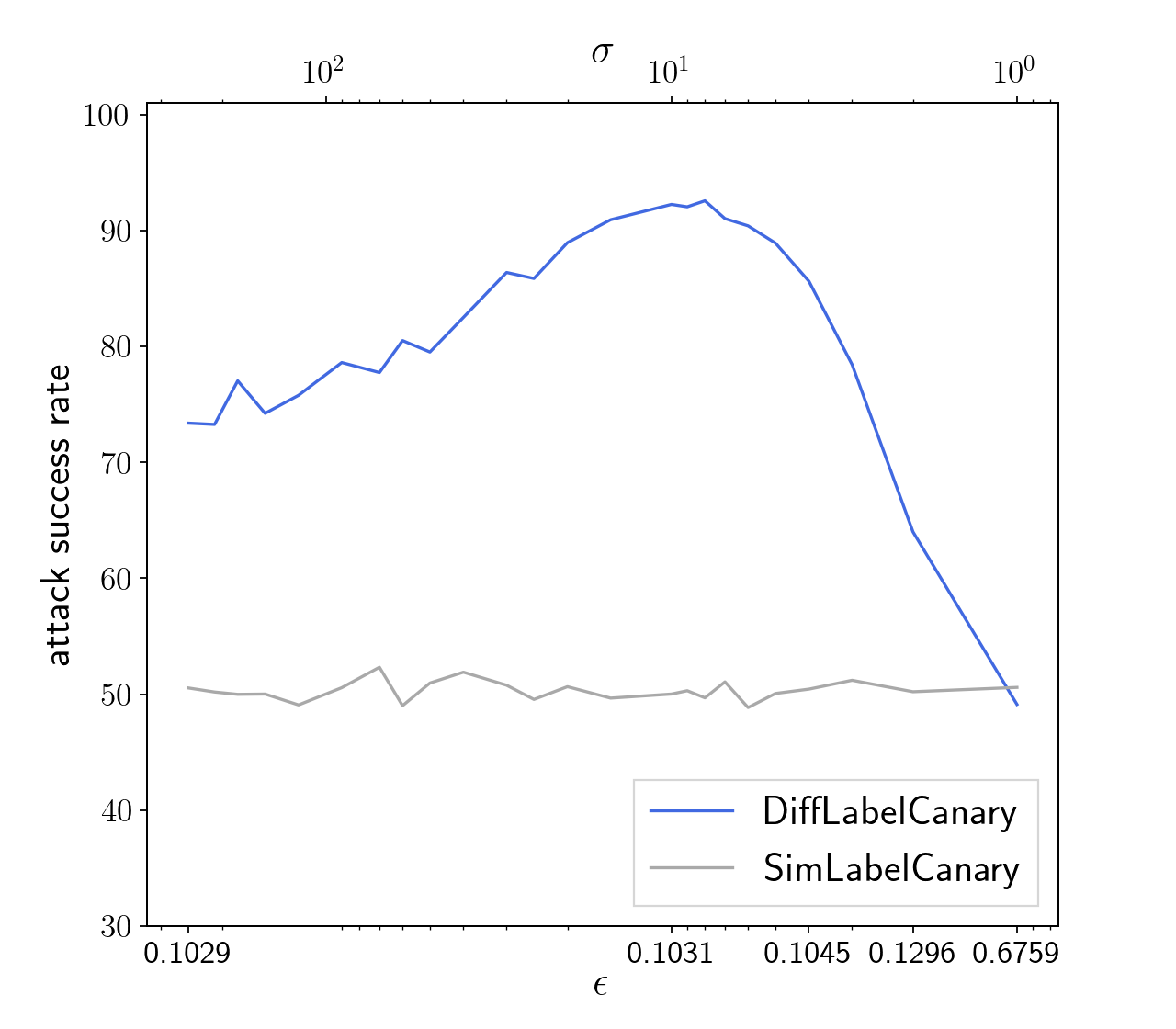}
  \hfill
  \caption{
  FP attack success rate  against DP-SGD on MNIST model using the Opacus~\cite{opacus}.
  $\difflabel$: attacker is distinguishing between a batch where all records have identical labels in the range $[1,9]$ and a batch
  that has same records as well as an image for label 0. $\samelabel$: batches where all records are from $[0,9]$
  and differ in one random record.
  Since the baseline attack is 50\%, the plot shows that FP attack is successful on $\difflabel$ where the canary record
  comes from the same distribution as MNIST data but different from other records in the batch. 
  Here, $\delta = 10^{-5}$.
  }
  \label{fig:mnist_acc_succ}
\end{figure}

\subsubsection{Attack Results}

The attack on DP-SGD follows the same procedure as the attack
in Section~\ref{float} since it uses $\gausspytorch$.
Note that the attack against DP-SGD naturally reveals sequential (cached)
samples from an implementation of the normal distribution
since the attacker observes application of noise to gradients of all $d$ parameters of the model,
i.e., to~$d$ computations that all use independent noise draws.

The adversary calculates $f(B)$ and $f(B')$ since we assume the attacker knows
the records in $B$ and $B'$ and is trying to determine the presence of the canary.
Since there are 26,010 gradients in $y$,
we evaluate the attack on each to see if it lies in support of $f(B)$ or $f(B')$. 
We extract $s_{1}$ and $s_{2}$, $s_{1}'$ and $s_2'$ from $y-f(B)$ and $y-f(B')$ respectively for each.
We then search for their neighboring values, and check whether any of them support
gradients in $y$. We say $y$ is in support of 
$f(B)$ if there are more gradients that support $f(B)$ than~$f(B')$ and similarly for $f(B')$.

The experimental results are presented in Figure~\ref{fig:mnist_acc_succ}.
For $\difflabel$ the attacker's succes rates is always better than a random guess for 
$\epsilon<0.68$.
Similar to previous results in this section, the attack success rate is much higher than 
the theoretical $(\epsilon,\delta)$-DP bound on failure of $\delta=10^{-5}$ would suggest.

We observe that the adversary cannot distinguish $f(B)$ and $f(B')$
when $\samelabel$ is used, that is, when the canary is similar to all the other records in the batch.
The reason is that the gradients for $B'$ and $B$ are close to each other and
hence, even when the noise is added the two stay relatively close to each other,
hence, the range of ``attackable'' floating-point they land on is similar.
On the other hand, for $\difflabel$ the canary record has a very different distribution
from records in $B$, thus the gradients are further apart which shifts them
into ranges of varying number of floating points.
It is important to note that the difference between
the gradients (i.e., sensitivity)
is protected by the DP mechanism based on a theoretical normal distribution using real values.
However, this difference is large enough that once the noise is added,
the noisy values are also shifted to ranges of floating points where one has more
attackable values than the other.
We observe a relative increase in gradient norm of 1.34
when using $\difflabel$ compared to  $\samelabel$.

We also observe that the success rate increases
when $\sigma$ increases, and correspondingly $\epsilon$ decreases.
This is counter-intuitive as the magnitude of noise increases as $\epsilon$ decreases. 
The same observation was made by Mironov for the Laplace distribution.
The reason again relates to the floating-point range in which noisy gradients land.
We also note that compared to attacks in~\ref{float},
the attacker has more chances to observe values in support of one dataset than the other,
since it has $d$ gradients to attack as opposed to one result.
\section{Discrete and Approximate Distribution Sampling}
\label{sec:timingtheory}
Discrete distributions aim to avoid privacy leakage from floating-point representation,
while retaining the privacy and utility properties of their continuous counterparts.
In this section we show that na{\"i}ve implementation of such discrete distributions suffers from 
a timing side channel attack: by measuring the time the sampling algorithm takes to draw noise from
such a distribution, the adversary is able to determine the magnitude of this noise
and, hence, invalidate the guarantees of differential privacy.

\paragraph{Discrete Laplace and Gaussian}
Canonne~\textit{et al.}~\cite{discrete}
study the discrete Gaussian mechanism and its properties.
They demonstrate that it provides the same level of privacy and utility as the continuous Gaussian.
Their sampler for Laplace and Gaussian uses the geometric distribution
where the corresponding samples preserve the magnitude of the noise drawn from the geometric distribution.
Unfortunately, the running time of this geometric distribution sampler,
if not implemented carefully,
reveals the magnitude of its noise.

Canonne~\textit{et al.}~\cite{discrete} describe sampling from geometric distributions (Algorithm 2 in their paper)
using Bernoulli samples.
Recall that the geometric distribution measures the probability of taking $n$ 
Bernoulli trials to obtain a first success.
Their pseudo-code to simulate this process proceeds as follows.
It samples the Bernoulli distribution until the first success while incrementing a counter of failures.
Once the success is observed, the counter is returned as a sample $n$ of a geometric distribution.
Hence, the number of times Bernoulli sampler is invoked linearly correlates with the magnitude
of the sample~$n$.
This is the source of the timing side-channel where the time
is correlated with a secretly drawn value.

The Laplace distribution~\cite{discrete} is based on a linear transformation
of the sample drawn from the geometric distribution, preserving its magnitude.
In turn, the discrete Gaussian mechanism with standard deviation of $\sigma$ in~\cite{discrete} uses the Laplace mechanism with parameter $\lfloor \sigma \rfloor + 1$.
Laplace noise is returned as-is via rejection sampling with a carefully chosen probability that produces the exact discrete distribution.
However, since Laplace noise is returned as-is, the Gaussian sample has the same magnitude as Laplace and hence as the geometric distribution.

In summary, the time it takes to run discrete Laplace or discrete Gaussian sampling
is correlated with the magnitude of the sample they return, and hence the noise
they add to their respective DP mechanisms.
Though the authors state that their algorithms may suffer from timing attacks,
they attribute them to rejection sampling noting that this ``reveals nothing about the accepted candidate''.
However, as we argue above and experimentally show in the next section,
the subroutine used to draw from the geometric distribution is the one that creates the timing side channel.

\paragraph{Approximate Laplace}
In parallel to the work by Canonne~\textit{et al.}, the Differential Privacy Team at Google~\cite{google_gauss} proposed
an algorithm for approximate Laplace in their report of the library implementation
for differential privacy~\cite{google-dp}.
Their sampler makes a draw from the geometric distribution
and then scales it using a resolution parameter based on~$\epsilon$.
The paper does not specify how the geometric distribution is implemented.
Upon examination of the code in the library~\cite{google-dp},
we observed that geometric sampling is not based on drawing Bernoulli samples
as in~\cite{discrete}.
However, its runtime still linearly depends on the value being sampled
and hence also suffers from a timing side-channel.
Specifically, the implementation of~\cite{google_gauss},
performs a binary search, where the distribution support region is split proportional
to probability mass and is guided by a sequence of uniform random values.
Since the search is longer for events with smaller probability, the time
to ``find'' larger values
in the case of drawing geometric random variables takes longer.
As a result this also creates a timing side channel that reveals the magnitude of the drawn noise,
even though conceptually the technique for drawing from the geometric distribution 
is different from~\cite{discrete}.

\section{Experiments: Timing Attacks on Discrete Distributions}
\label{sec:approxmatch}

We evaluate the discrete Laplace and Gaussian using
the implementation in~\cite{discrete-imp} that accompanied the work by Canonne~\textit{et al.}~\cite{discrete} and the Laplace implementation from
Google~\cite{google_gauss}, referred to as implementations \RNum{1} and~\RNum{2} respectively
(see disclosure in~Section~\ref{sec:intro}).
We conduct two sets of experiments to show that
{1)}~both implementations are amenable to timing attacks;
{2)}~a DP algorithm that uses these implementations, 
as a consequence, is also amenable to a timing attack.

\subsection{Experimental Setup}
We run the discrete samplers on a single core of an Intel Xeon Platinum 8180M, 
which runs a 64-bit Ubuntu Linux 16.04.1 with kernel version 4.15.0-142. 
There are no other running processes on this core, 
so the interference on timing measurement is minimized. 
We measure the overall time of the sampling algorithm on invocation and exit 
with nano second precision using \texttt{time.process\_time\_ns()} for Implementation~\RNum{1} 
written in Python and \texttt{System.nanoTime()} for Implementation~\RNum{2} written in Java.

\subsection{{Timing of Discrete Samplers}}
\label{sec:timsampleexp}
We first measure the time it takes to generate the noise
from each implementation and average it over more than 1 million trials
for each sampled value in a truncated region.
Since both Gaussian and Laplace are symmetrical distributions, the time
it takes to generate positive and negative noise is also symmetric.
In the implementation, the sign is determined independently from the (positive) geometric noise magnitude.

\begin{figure*}
  \centering
  \hfill
  \subfigure[]{\includegraphics[width=8.7cm]{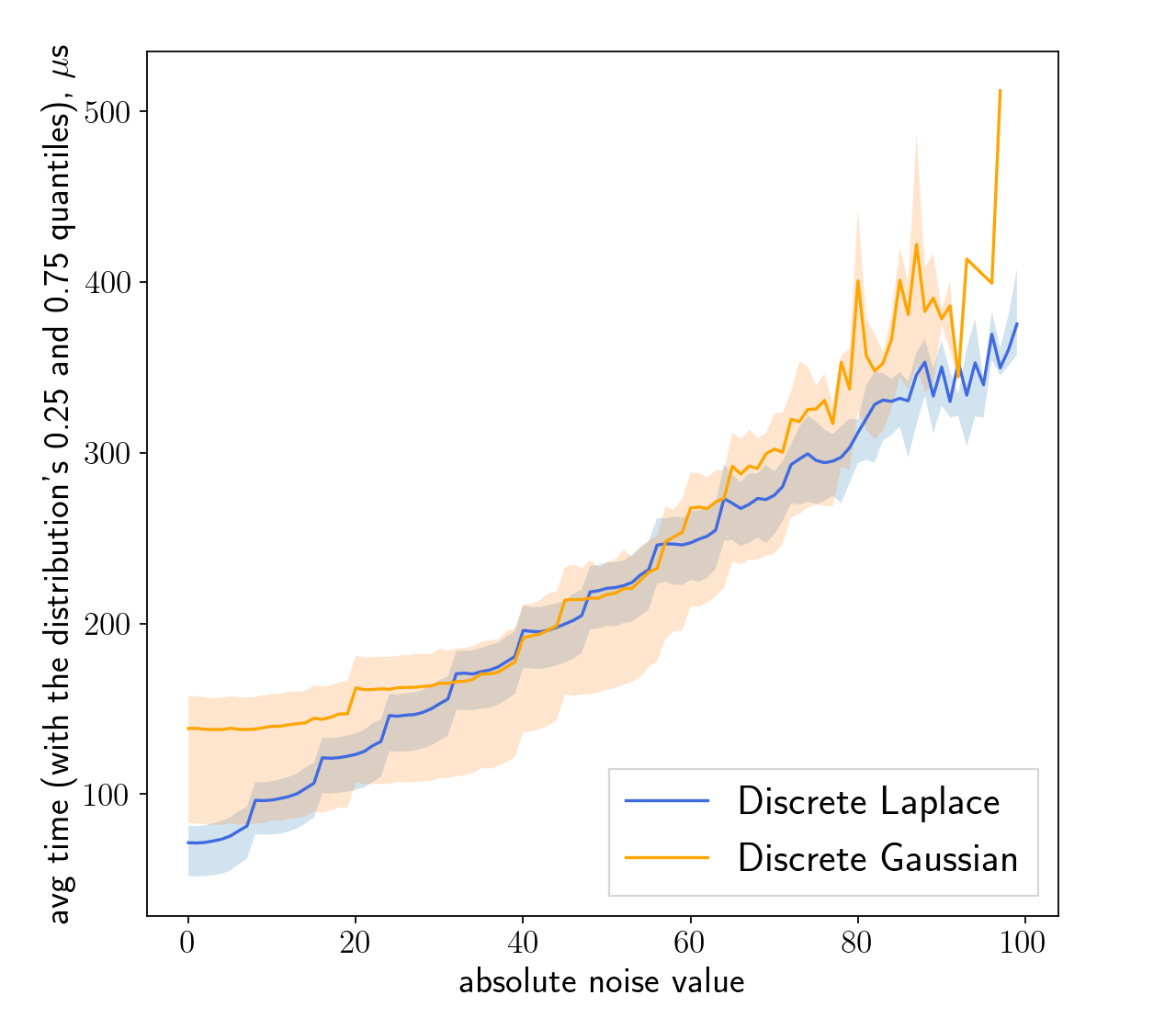}}
  \hfill
  \subfigure[]{\includegraphics[width=8.7cm]{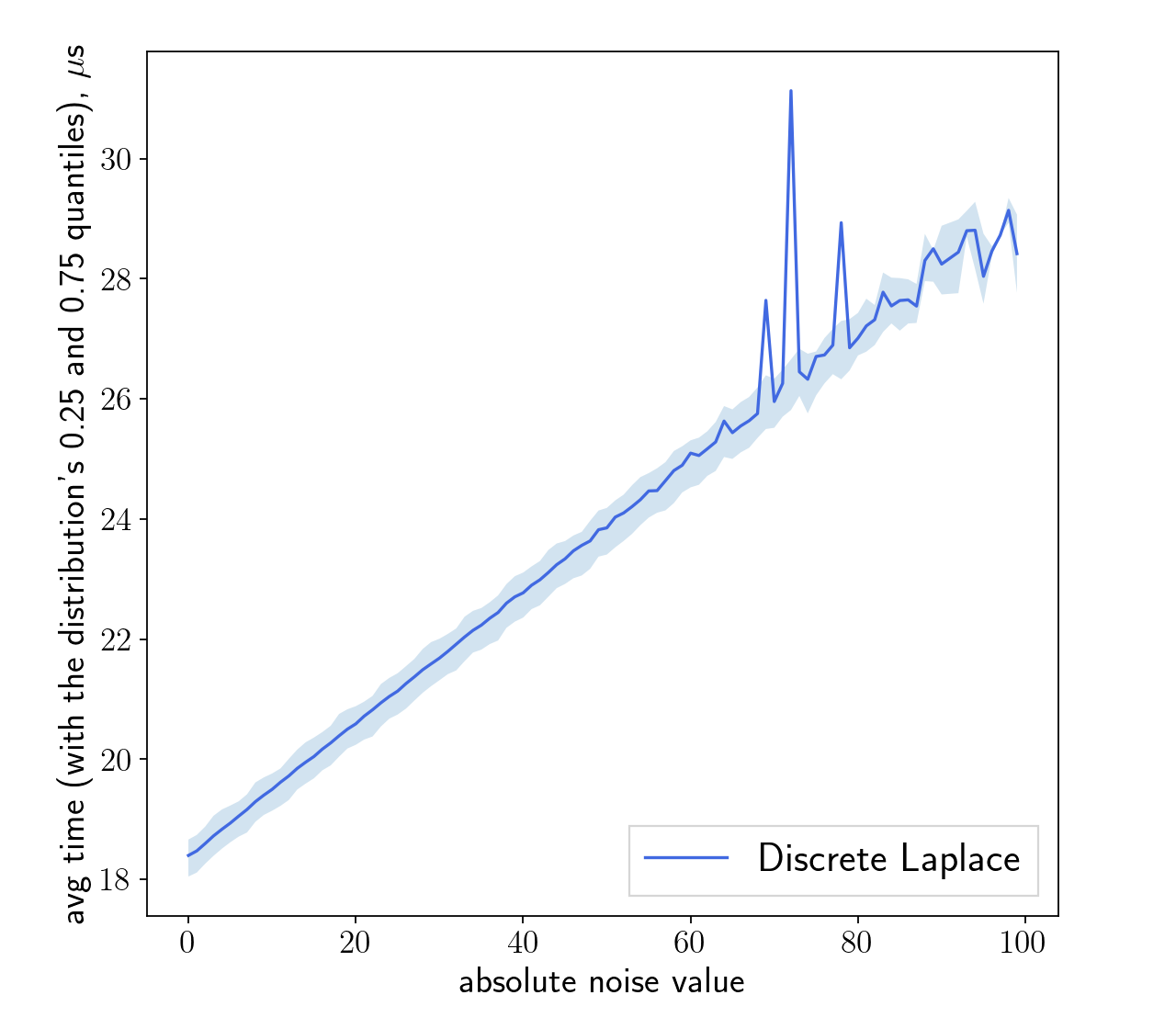}}
  \hfill
  \caption{Average time in $\mu$s (with the distribution's 0.25 and 0.75 quantiles)
  to generate absolute noise using two implementations of discrete distributions:
  (a) Discrete Laplace and Gaussian from Implementation \RNum{1}~\cite{discrete-imp} (avg.~over 10 million trials).
  (b) Discrete Laplace from Implementation \RNum{2}~\cite{google-dp} (avg.~over 20 million trials).
  The plots show a linear relationship between the absolute noise value and time it takes to generate it.
  Gaussian~\RNum{1} uses $\sigma=19$, Laplace~\RNum{1} uses $\lambda=8$ and
  Laplace~\RNum{2} uses $\lambda=\frac{8}{\ln{3}}$.}
  \label{fig:timing}
\end{figure*}

In Figure~\ref{fig:timing} we plot the average time it takes to draw absolute values of the noise.
We used $\sigma=19$ for Gaussian~\RNum{1}, 
$\lambda=8$ for Laplace~\RNum{1}, and $\lambda=\frac{8}{\ln{3}}$ for Laplace~\RNum{2}.
We observe that the absolute magnitude of noise has a positive
linear relationship with time to generate noise from all implementations.

Based on the above relationship between noise magnitude and generation time,
we implement our attack as follows.
The attacker computes average time $t_i$ to generate absolute values of integer noise $i \in [0,9]$.
It then measures the time $t_j$ it takes to generate an unknown noise $j$ sample and chooses $i$ that has the closest time to $t_j$
as its guess:
\newcommand{\guess}{\mathsf{guess}}
\begin{equation}
\label{guess}
j_\guess = {\arg\min}_{i \in [0,9]}(|t_j-t_{i}|)
\end{equation}

We ran 100,000 trials for each sampler
to evaluate the accuracy of our timing attack.
We evaluate accuracy in two standards: exact match and approximate match within $\pm 1$ from the correct value:
\begin{itemize}
  \item exact match: $|j| = j_\guess$
  \item approximate match: $-1 \le j_\guess-|j|\le1$
\end{itemize}

The attacker can use exact match as follows.
Recall that the attacker, given a DP output $y$ and $j_\guess$ where $y= f(\cd) + j$
is trying to guess the unprotected value of $f(\cd)$.
Let the co-domain of $f$ have integer support $[-X, X]$, 
known to the adversary as it knows $f$ and domain of $f$, $\mathcal{D}$.
For exact match though our attack cannot guess whether the noise is positive or negative,
this still harms differential privacy. Specifically, 
it reduces the original guess of the attack on the noise value
from $1/(2X +1)$ to $1/2$.

For the approximate guess, the attacker knows that $j$
could have been one of six values:
$$\pm (j_\guess -1) , \pm j_\guess, \pm (j_\guess +1)$$
This allows it to determine that $f(D)$ must be either $y \pm  (j_\guess -1)$,
$y \pm j_\guess$ or $y \pm  (j_\guess +1)$. Hence, its guess
is reduced from $1/(2X +1)$ to $1/6$.
As an example, suppose the attacker is trying to distinguish
between $f(\cd) = 30$ and $f(\cd') = 60$. If it observes,
$y=20$ and $j_\guess=9$. It knows that $f$ must have been 
computed on $\cd$ and not $\cd'$.

  \begin{table}[t]
\normalsize
  \centering
  \begin{tabular}{cccc}
  \hline
  Implementation                       & Parameters                                                                              & Match    & Accuracy \\ \hline
  \multirow{4}{*}{Gaussian \RNum{1}~\cite{discrete-imp}} & \multirow{2}{*}{$\sigma=2$}                                                             & exact       & 24.4\%   \\
                                       &                                                                                         & approximate & 56.2\%   \\ \cline{2-4} 
                                       & \multirow{2}{*}{$\sigma=4$}                                                             & exact       & 13.6\%   \\
                                       &                                                                                         & approximate & 39.9\%   \\ \hline
  \multirow{4}{*}{Laplace \RNum{1}~\cite{discrete-imp}}  & \multirow{2}{*}{$\lambda=1$}                                                             & exact       & 42.1\%   \\
                                       &                                                                                         & approximate & 84.2\%   \\ \cline{2-4} 
                                       & \multirow{2}{*}{$\lambda=3$}                                                             & exact       & 24.7\%   \\
                                       &                                                                                         & approximate & 61.5\%   \\ \hline
  \multirow{6}{*}{Laplace \RNum{2}~\cite{google-dp}}  & \multirow{2}{*}{$\lambda=\frac{1}{\ln{3}}$}                                               & exact       & 42.0\%   \\
                                       &                                                                                         & approximate & 89.0\%   \\ \cline{2-4} 
                                       & \multirow{2}{*}{$\lambda=\frac{3}{\ln{3}}$}                                             & exact       & 17.0\%   \\
                                       &                                                                                         & approximate & 44.0\%   \\ \cline{2-4} 
                                       & \multirow{2}{*}{$\lambda=\frac{1}{\ln{2}}$}                                             & exact       & 32.0\%   \\
                                       &                                                                                         & approximate & 71.0\%   \\ \hline
  \end{tabular}
  \caption{Success rates of timing side channel attacks against implementations of discrete distributions.
  The baseline accuracy based on a random guess for exact and approximate match is $10\%$ and $33\%$, respectively.
  We observe that the attacks are always well above the baseline attack accuracy.}
  \label{tab:appdiscrete}
  \end{table}

The results are summarized in Table~\ref{tab:appdiscrete}
where we measure the accuracy for samples
in the range $[-9,9]$ for several parameters where $\epsilon$ ranges between 0.5 and 2 and $\delta = 10^{-5}$.
Since we evaluate the absolute values of the sample,
the baseline accuracy based on a random guess for exact and approximate match is $10\%$ and $33\%$, respectively.
We observe that the attacks are always well above the baseline accuracy.
This indicates that if an adversary can observe how long the sampling algorithm takes to 
generate noise used in a DP mechanism, then it can guess the relative magnitude of this noise for samplers based
on geometric noise generation.
  
 For Implementation~\RNum{1},
 the discrete Laplace mechanism is more vulnerable to timing attacks than the discrete Gaussian.
 This is likely due to rejection sampling that Gaussian implementation adds to the process.
 Rejection sampling adds stochasticity to which Laplace sample, among several that are drawn,
 is returned.
 For example, the attacker cannot distinguish between the following two executions:
 \begin{enumerate}
   \item In the first execution, a large Laplace noise value is generated but then rejected,
   while a small Laplace noise value generated next is returned as a result.
   \item 
   In the second execution, the opposite happens where a small Laplace value is rejected first
   and then a larger Laplace value is generated and returned as a result.
 \end{enumerate}
 Execution times will be similar for both cases.
 Indeed, the authors of~\cite{google_gauss} also proposes a discrete Gaussian based on Binomial
 sampling and rejection sampling and their implementation~\cite{google-dp}
 is not amenable to our attack.

The success rate of our attack decreases with increasing $\sigma$ and $\lambda$.
We note that with higher parameters, larger noise values are more likely,
while the attack is more successful for smaller values, hence,
overall accuracy decreases.
For example, with $\lambda=3$
for Laplace~\RNum{1}, the accuracy  for
$i\in[0,4]$ is 27.1\%, and accuracy for $i\in[5,9]$ is 12.5\%.

For Laplace~\RNum{2} we observe a similar trend as for Laplace~\RNum{1}
even though the geometric distribution is sampled using a different procedure described in
the previous section.

\subsection{Timing Attack on Private Sum}
\label{sec:e2e}

Based on the results of the previous section, we conduct a timing attack on real data using
the German Credit Dataset~\cite{german-credits} used in Section~\ref{sec:privatecountFP}.
We model the setting where the private sum of the credit attribute of the dataset
is computed (e.g.,
an analyst queries a DP-protected database as in the DB setting in Section~\ref{sec:threat}).
The attacker is trying to guess the non-private sum using query's response and the time it takes for the query to return.
We put a limit that each
individual can have at most 5000 credits (which determines the
sensitivity of the query).
We use private sum computation which mirrors private count in Section~\ref{sec:privatecountFP}
except $f$ is a sum and $s$ is sampled from Laplace~\RNum{2}~\cite{google-dp}.
The neighboring datasets $D$ and $D'$ 
that our attack tries to distinguish
differ on a single record whose credit is 5000 in $D$
and 0 in $D'$.
We note that the query is different from that in Section~\ref{sec:privatecountFP} since the timing
attack is not as effective for queries with small sensitivity, while the FP attack is effective against
queries that return values close to 0.

To conduct the attack, we first collect timing data of
private sum for different noise magnitudes.
Similar to experiments in the previous subsection, we observe a
linear relationship between noise magnitude and time
cost albeit now this time includes computing the sum and sampling
(Figure~\ref{fig:appprivate_sum} in Appendix).

Our attack proceeds by measuring the time $t$ of a DP algorithm to complete.
The attacker then uses
$t$ and the output $y$ it receives
to determine if it was $D$ or $D'$ used in the computation.
Note that the noise magnitude
should be $s = |y-\sum_{D} x|$ for $D$ and $s' = |y-\sum_{D'} x|$ for $D'$.

The attacker makes a guess on
the magnitude of the noise using the timing data it has collected above
and Equation~\ref{guess}. Let $s_g$ be its guess.
It then compares $s_g$ with $s$ and $s'$ and chooses the closest one as its
guess.
That is, it chooses $D$ if $|s-s_g | <|s' - s_g|$ and $D'$ otherwise.

\begin{figure}[t]
  \includegraphics[width=0.45\textwidth]{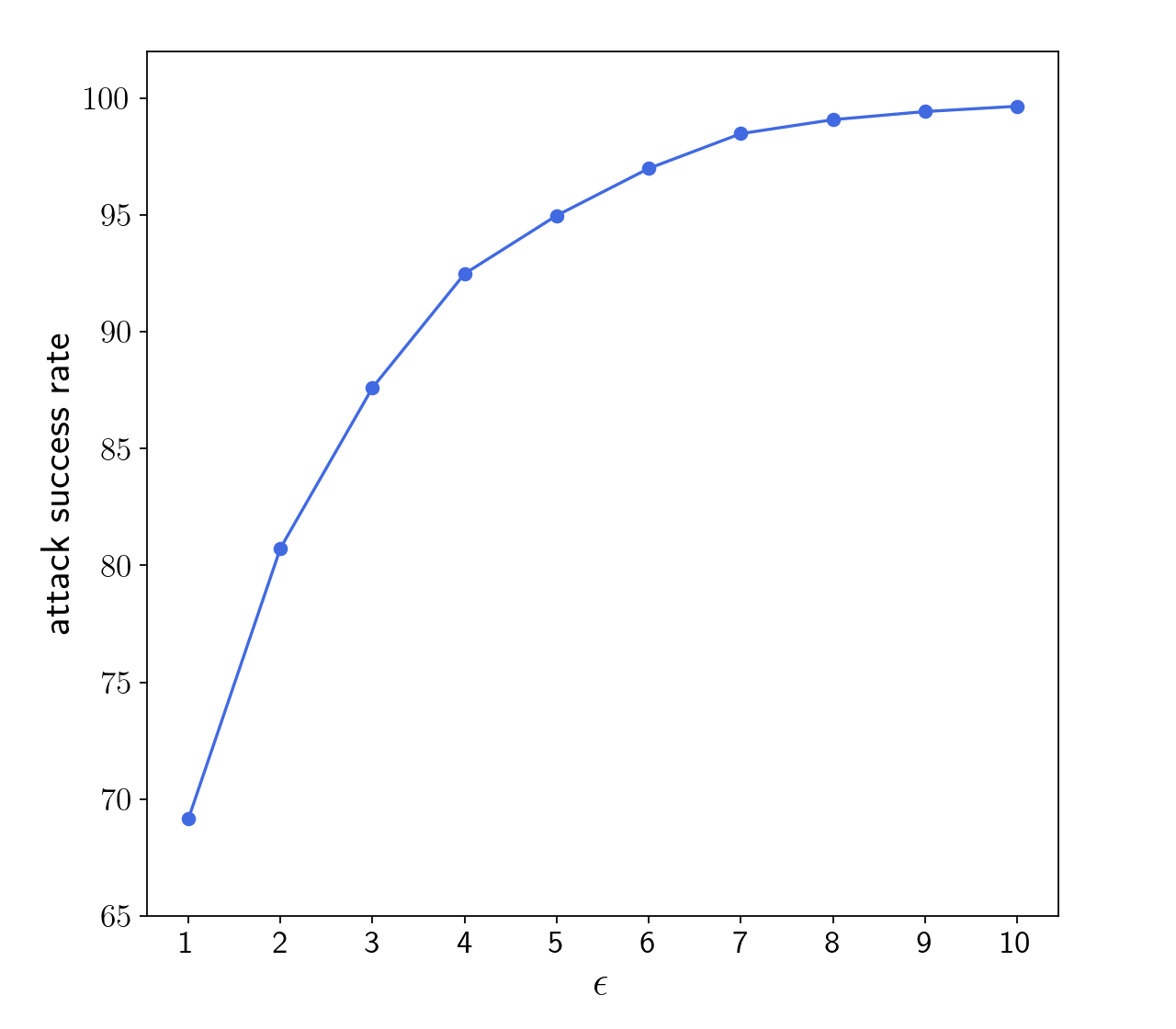}
  \centering
  \caption{Attack on private sum of the credit attribute of the German Credit 
  Dataset~\cite{german-credits}, with Laplace~\RNum{2}~\cite{google-dp}.
  The attack success rate under different privacy budget 
  $\epsilon\in[1,10]$ and sensitivity $\Delta=5000$ (measured over 1 million
  trials for each $\epsilon$).
  Success means the private sum
  created from $D$ is successfully concluded to be in support
  of $D$ and not $D'$.}
  \label{fig:private_sum}
\end{figure}

We plot the attack results in Figure~\ref{fig:private_sum}
for $\epsilon$ in the range $[1,10]$.
We observe that attack success rate increases with higher $\epsilon$, 
with success rate of 69.15\% for $\epsilon=1$.
Our intuition for the above trend is due to the range of noise
in which the attacker needs to make its guess. That is,
for smaller noise scale (i.e., high $\epsilon$),
the output noise range is small as well and hence
attacker has less number of noises to assign observed time to.
For example, with $\Delta=5000$, noise magnitudes are mainly distributed in
$[0,15000]$ when $\epsilon=1$, while noise magnitudes 
are mainly distributed in $[0,2500]$ when $\epsilon=10$.
In Appendix~\ref{app:probattack} we  compare the success rate of the timing attack with the success rate of an attacker that
uses knowledge of the probability distribution used by a DP mechanism.

\medskip

The timing attack has two limitations. First, it assumes that the time (and its variance)
to compute $f$ is not much larger than that of sampling, as otherwise the microseconds difference
may not be observable. Second, the attack works
for one-dimensional functions $f$ as the attacker can measure the time of a single noise sample.
Hence, the attack will not be as successful if the attacker were to observe the time it takes to
draw multiple samples, as is the case for DP-SGD.

\section{Mitigation Strategies}
\label{sec:mitigations}

We discuss mitigation strategies for both of our attacks.

\subsection{Defenses Against Floating-Point Attacks}
\label{sec:deffp}

Mironov~\cite{mironov} proposed the snapping mechanism to alleviate the FP attack
by carefully truncating and rounding an output of a DP mechanism that was implemented using floating points.
However, the privacy and utility of the overall mechanism decreases~\cite{google_gauss,discrete}.

Our attack {against the Box-Muller and polar methods} assumes 
that the attacker observes two {of their} samples
({recall that the Ziggurat method generates only one sample}).
That is, the attacker gets access to the second (cached) value 
(e.g., when a query returns an answer to a $d$-dimensional query
such as a histogram or ML model parameters).
{W}ithout the second value
in Equation{~\eqref{eq:fpattack} the attacker has to resort 
to checking all possible values in a brute-force manner}.
A potential mitigation
is therefore to generate new samples on each call
and disregard the second value.

We also observe that the implementation of DP-SGD adds noise to the batch directly.
Instead it could potentially add several samples of noise
and then average the result, since an average of Gaussian noise is still Gaussian.
This would make it harder for an adversary to extract sample-level values needed for the FP attack
as described in~Section~\ref{sec:fpattack}. 
However, this heuristic may still be susceptible to attacks
as it also uses floating-point representation.

Discrete distributions~\cite{discrete,google_gauss,DBLP:conf/innovations/BalcerV18} have been proposed 
as a mitigation against floating-point attacks since they avoid
floating points or bound their effect on privacy in $(\epsilon,\delta)$ parameters.
However, as we showed in the previous section, they may suffer
from other side-channel attacks and thus should also be carefully implemented.
Nevertheless in this section we evaluate them as a mitigation
for attacks in Section~\ref{sec:dpsgdattack} and measure their effect on accuracy
of DP-SGD.

We  implement DP-SGD with discrete Gaussian by discretizing 
the gradient $g\in\mathbb{R}^d$ as described in~\cite{discrete-fl}
who use discrete Gaussian for training models in a Federated Learning setting.
Appendix~\ref{app:discmitig} provides further details.

We use the same model and MNIST dataset as in Section~\ref{sec:dpsgdattack}
to evaluate the performance of models trained with discrete Gaussian
for a discretization parameter $\gamma\in\{10^{-1},10^{-2},10^{-3}\}$ and
privacy budget $\epsilon\in(0.1,10]$.
We use the implementation of discrete Gaussian \RNum{1}~\cite{discrete-imp}.

\begin{figure}[t!]
  \includegraphics[width=0.45\textwidth]{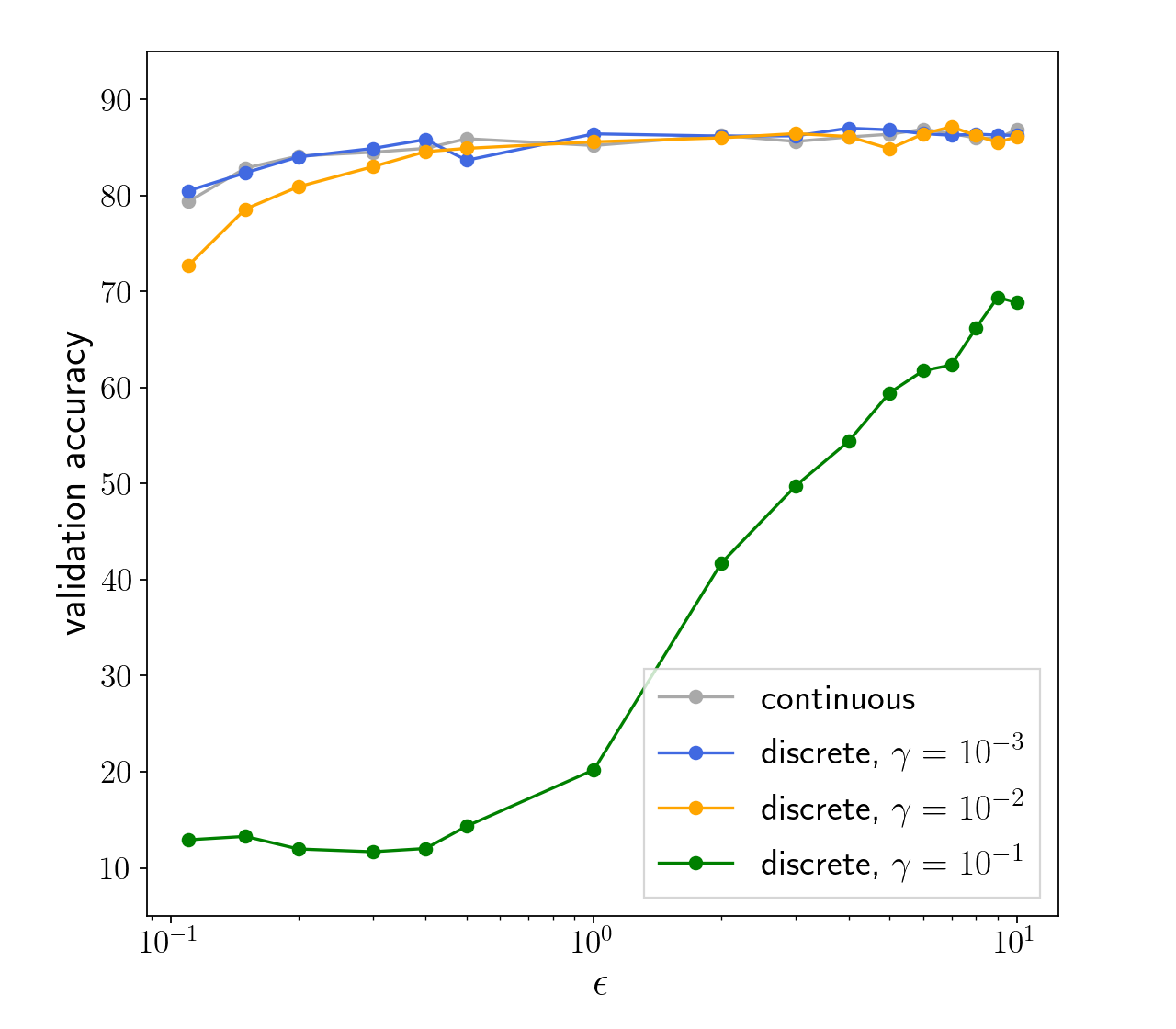}
  \centering
  \caption{Accuracy of the ML model on MNIST dataset, trained with
  DP-SGD with continuous and discrete Gaussian samplers with
  discretization parameters $\gamma\in\{10^{-1}, 10^{-2},10^{-3}\}$
  for $\epsilon\in(0.1,10]$.}
  \label{fig:mnist_discrete_eps}
\end{figure}

We 
compare accuracy of the models with discrete and continuous implementations
in Figure~\ref{fig:mnist_discrete_eps}.
We observe that the performance of models
trained with discrete Gaussian matches the
performance of models trained with continuous Gaussian, given small enough 
$\gamma$, such as $10^{-3}$.
Smaller $\gamma$ allows the discretization to be done on a finer 
grid $\gamma\mathbb{Z}$, thus gradient calculation is not affected by
rounding.

\subsection{Defenses Against Timing Attacks} 
\label{sec:deftiming}

The most effective mitigation against timing attacks
is to ensure constant time execution. 
Application-independent approaches to do so
include compiler-based code 
transformations~\cite{10.1145/2086696.2086702,10.1145/3213846.3213851}
while other generic defense strategies are based on padding
either by padding execution time to a constant time or adding a random
delay~\cite{10.1145/1866307.1866341}.

In our setting, one could choose a sufficiently large time threshold and run noise generation mechanism
until then, independent of the noise being drawn 
and the time that takes to produce it.
If an attacker is stronger than that considered in this paper and can perform microarchitectural observations,
then execution of any padded code has to be made secret-independent.
For example, if the attacker can measure the time of memory accesses, 
the counter of the number of trials needs to be accessed regardless of whether heads or tails was drawn in a Bernoulli trial.
The downside of padding is efficiency as
all execution would take maximum time.
The failure
to draw noise within the threshold time can then be accounted for in~$\delta$, the failure probability of DP~\cite{discrete}.
An alternative is to use a truncated version
of the geometric distribution~\cite{10.5555/2028067.2028100} in order to avoid values that
are impossible to sample on a finite computer.

We evaluate padding as a mitigation against our attack on private sum in Section~\ref{sec:e2e}.
Padding can be implemented via two approaches. First,
after a successful trial is encountered, we record its trial number
and continue drawing ``blank'' Bernoulli samples,
till the maximum noise threshold is reached.
In the second approach, once a successful trial is encountered, a time delay can be added to reach
some maximum time threshold.
We choose the maximum threshold to be 
41$\mu$ since we observed that 
99.5\% of noise magnitudes are distributed in $[0, 2677]$,
with average and maximum execution time of 37.1$\mu$
and 40.3$\mu$, respectively,
when $\epsilon=10$ and $\Delta=5000$.
We note that these estimates
are not dependent on the data but only on the sampler and the parameters of the noise.
Padding until a time threshold increases the total execution by 10.5\%.
For medium number of queries this is an acceptable overhead given the small 
magnitude of the overall time (in $\mu s$). Note that these estimates will be different for other
implementations.

As another mitigation strategy we propose a technique based on batching and caching.
The method generates $k$ random values {offline} and saves them.
It returns {one} value {for each call to the distribution function}.
It proceeds this way until all cached values are used and then restarts 
the process by generating the next $k$~samples.
Samples could be generated
online and shuffled to disconnect noise from their timing, as suggested in~\cite{9358267}.
However, the attacker may still measure the range of possibles times.

\medskip

In summary, discrete distribution sampling implemented in constant time or 
where the timing of sampling is not observable  (i.e., generated offline)
appears to be the best approach for defending against attacks discussed in this paper.
\section{Related Work}
\label{sec:related}
The implementation of differential privacy via Laplace mechanism has been
demonstrated to be flawed, due to finite-precision representations of floating-point values~\cite{mironov}. In this paper, we demonstrate
that implementations of the Gaussian mechanism suffer
from the same attack, with adjustments to the attack process.
In~\cite{DBLP:journals/corr/GazeauMP13} the authors demonstrate
an attack against DP mechanisms that
use finite-precision representations, and propose a mitigation strategy.
Recently Ilvento~\cite{10.1145/3372297.3417269}
has explored practical considerations and pitfalls of implementing the exponential mechanism
using floating-point arithmetic. They show that
such implementations are also susceptible to attacks and propose
a solution using a base-2 version of the exponential mechanism.

Timing attacks against DP mechanisms have been explored in~\cite{10.5555/2028067.2028100}
and~\cite{DBLP:conf/sp/AndryscoKMJLS15}. However, they differ from the attacks described in this paper
as they do not exploit the timing discrepancies introduced by distribution sampling.
In~\cite{10.5555/2028067.2028100}, the authors observe that the mechanism implementation may suffer from timing attacks
(e.g.,~because it performs conditional execution based on a secret).
As a mitigation authors propose constant time execution for the mechanism, without consideration of noise generation. 
On the other hand, Andrysco~\textit{et al.}~\cite{DBLP:conf/sp/AndryscoKMJLS15} 
exploit the difference in timing of floating-point arithmetic operations
(e.g.,~multiplication by zero takes observably less time than multiplication by a non-zero value).
They also show that DP mechanisms {(including~\cite{10.5555/2028067.2028100})} 
are susceptible to information leakage by using floating-point instructions whose
running time depends on their operands.

Balcer and Vadhan~\cite{DBLP:conf/innovations/BalcerV18} outline shortcomings
of implementing differentially-private mechanisms on finite-precision computers,
including a discussion of floating-point representations and sampling from distributions with infinite support.
The authors propose a polynomial-time discrete method for answering approximate histograms.
Their method is based on a bounded (or truncated) geometric distribution. Though a full implementation is not provided,
the authors suggest that the distribution can be sampled via inverse transform sampling using binary search over the support range
using cumulative distribution function $F$ to guide the search.
That is, given a uniform random value $p$ sampled uniformly from $(0,1]$,
find smallest value $x$ from the support of the geometric distribution such that $F(x) \ge p$.
If performed na{\"i}vely, such an approach could reveal the magnitude of the noise
of the geometric distribution since the search will take longer for ``less likely'' values (i.e., larger $p$)
and would therefore suffer from the same timing channel as described in Section~\ref{sec:timingtheory}.
In a follow up work to ours, Ratliff and Vadhan~\cite{ccs:RatliffZ24} 
have proposed a DP framework to explicitly capture timing side channels.

In independent and parallel work~\cite{DBLP:journals/corr/abs-2107-10138},
the authors describe a theoretical attack against the Box-Muller method of sampling from the Gaussian distribution in a similar manner to our FP attack. However, they do not provide experiments
validating the attack's efficacy against existing implementations. 
The authors propose a mitigation strategy similar to the one mentioned in Section~\ref{sec:mitigations}
based on computing a Gaussian sample from multiple samples, and analyze its robustness. Their work does not consider timing attacks.

Several systems have been proposed to
estimate a lower bound on $\epsilon$
that is achieved by a given DP mechanism in practice by finding counterexamples that would violate the theoretical guarantee
(see~\cite{Bichsel2021DPSniperBD, bichsel2018dp, ding2018detecting} and references therein).
For example, by training a ML classifier to distinguish outputs produced by neighbouring inputs,
DP-Sniper~\cite{Bichsel2021DPSniperBD} can find a higher $\epsilon$ than the one proved in theory
for the ``naive'' implementation of Laplace sampler.

Stepping away from differential privacy,
Gaussian samplers have been also widely used in schemes
for  digital signatures, public key encryption, and key exchange based 
on lattice based cryptography~\cite{Follath+2015+1+23,10.1145/2535925,10.1007/978-3-642-40041-4_3}.
Such cryptographic primitives often rely on a multi-dimensional discrete Gaussian which is approximated by
a distribution that is statistically close to the desired distribution.
Several sampling mechanisms have been shown to suffer from cache-based side-channel 
attacks~\cite{10.1007/978-3-662-53140-2_16} (i.e., based on memory accesses of the underlying
algorithm)
and power analysis~\cite{9358267}.
Defenses based on constant-time execution~\cite{Howe2018OnPD,cryptoeprint:2017:259,8314133}
and shuffling~\cite{9358267} have been proposed to protect against some of these attacks
including timing.
Though some techniques proposed for hardening the code in this space
can be used for protecting samplers for DP (Section~\ref{sec:mitigations}),
direct use of such samplers for DP is not straightforward.
This follows from the observation that discrete variants in this area are
(only) statistically close to the desired discrete distribution. As a result, 
composition-based analysis developed for analyzing cumulative loss
of multiple mechanisms based on Gaussian noise~\cite{DBLP:journals/corr/DworkR16,privbook}
cannot be used directly.

\section{Conclusion}
In this paper we highlight two implementation flaws
of differentially private (DP) algorithms.
We first show that the widely used Gaussian mechanism
suffers from a floating-point (FP) attack against implementations of normal 
distribution sampling, similar to vulnerabilities of the Laplace mechanisms as demonstrated in 2011 by Mironov.
We empirically demonstrate that implementations in {\texttt{NumPy},~\texttt{PyTorch}} 
and \texttt{Go}, including those used in implementation of open-source DP libraries
are susceptible to the attack, hence violating their privacy guarantees.
Though some researchers have speculated that the Gaussian mechanism may be susceptible to FP attacks,
this is the first work to provide a comprehensive evaluation showing that it is feasible in practice.

In the second part of the paper we show that implementations of 
discrete Laplace and Gaussian mechanisms --- 
proposed as a remedy to the FP attack against their continuous 
counterparts --- are themselves vulnerable to another side-channel due to timing.
That is, we show that implementations of such discrete variants, including a DP library by Google,
exhibit the time that is correlated with the magnitude of the secret random noise.
Our work re-iterates the importance of careful implementation of DP mechanisms
in order to maintain their theoretical guarantees in practice.

\section*{Acknowledgment}
The authors are grateful to the anonymous reviewers for their feedback that
helped improve the paper.
This work was supported in part by a Facebook Research Grant and the joint CATCH MURI-AUSMURI.
The first author is supported by the University of Melbourne research scholarship (MRS) scheme.
We thank Thomas Steinke for pointing us to the Ziggurat method in Go
and Cl\'ement Canonne for insightful discussions on statistically-close Gaussian samplers.
We are grateful to Zachary Ratliff and Jacob Urick 
for sharing with us their comparison of a probability-based baseline with our timing attack,
which we investigate further in Appendix~\ref{app:probattack}.

\bibliographystyle{IEEEtran}
\bibliography{references}

\appendix

\begin{figure} [t]
  \includegraphics[width=8.7cm]{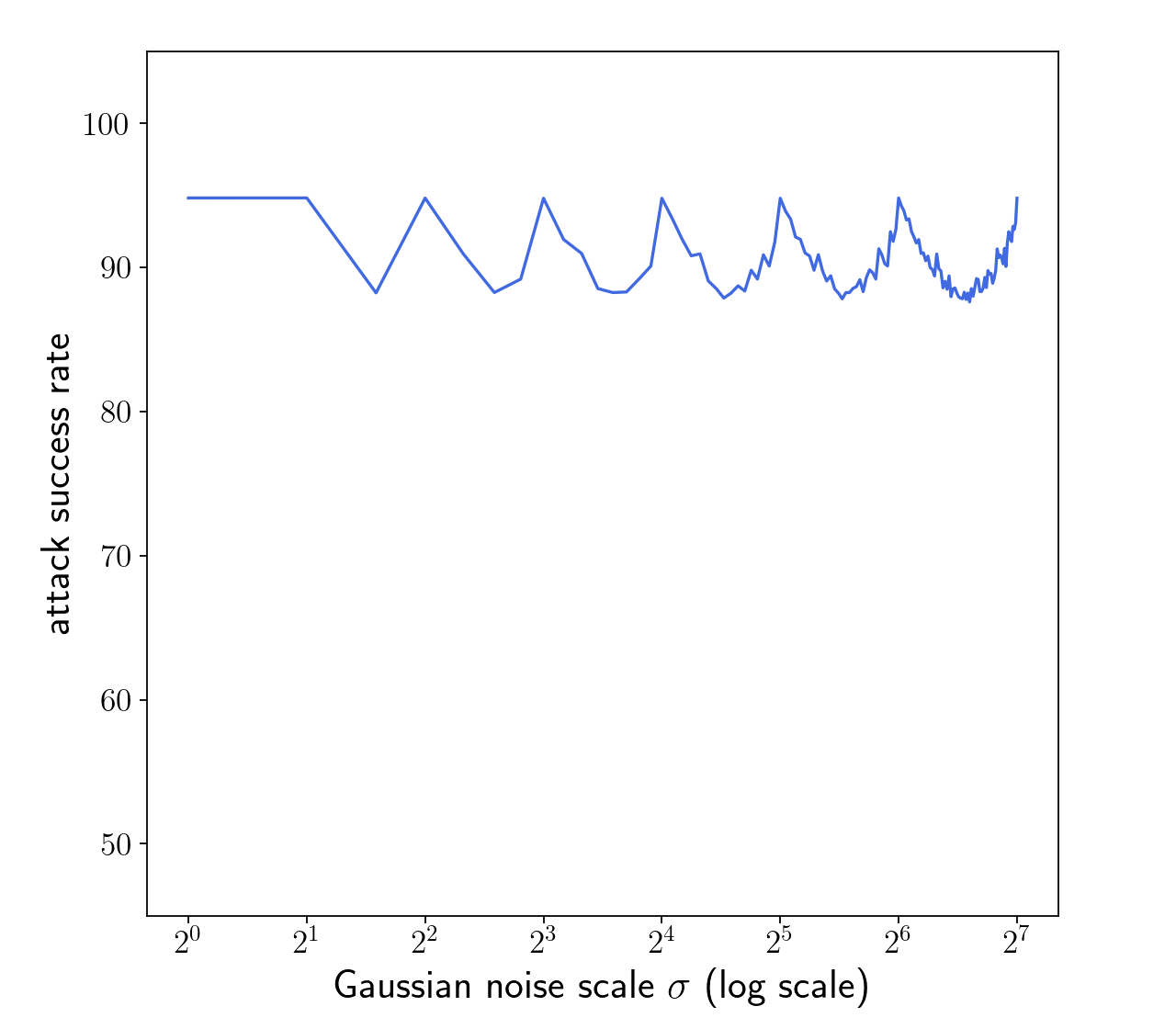}
  \caption{
    FP attack success rate on private count
    where the count is protected with the $\gaussgo$ Gaussian sampler
    across $\sigma \in [1,128]$ with fixed $\delta=10^{-5}$
    and function sensitivity $\Delta=1$.
    Baseline (random) attack success is~50\%.
    Success rate peaks at $\sigma \in \{2^0, 2^1, 2^2, 2^3, 2^4, 2^5, 2^6, 2^7\}$.
  }
  \label{fig:zig_pattern}
\end{figure}

\begin{figure} [t]
\includegraphics[width=8.7cm]{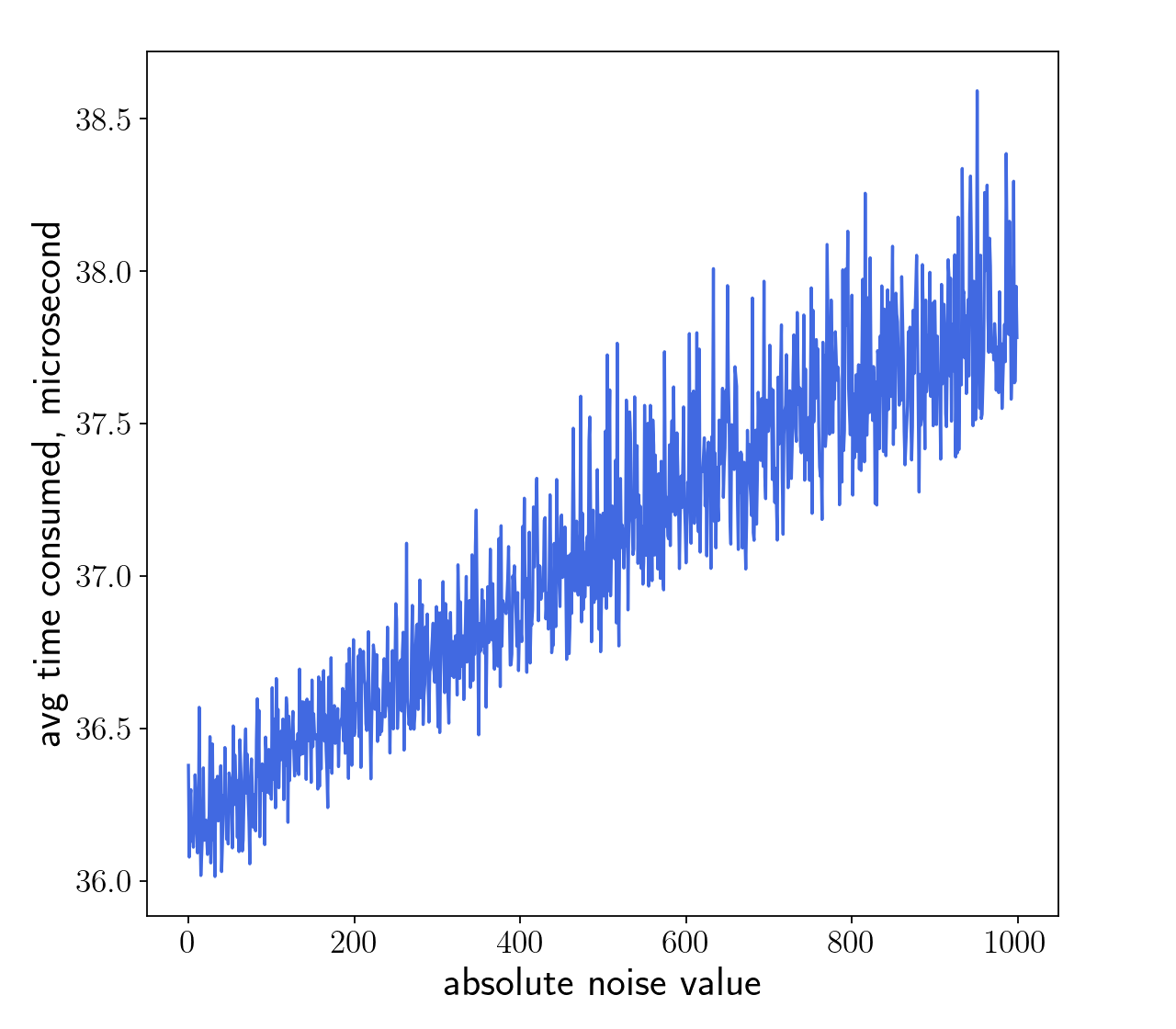}
\caption{
Execution time of private sum over
the credit attribute of the German Credit Dataset~\cite{german-credits}, with Laplace~\RNum{2}~\cite{google-dp},
over a range of noise magnitudes, 
with $\epsilon = 10$ and sensitivity $\Delta = 5000$ (avg. over 50 million trials).}
\label{fig:appprivate_sum}
\end{figure}

\subsection{Box-Muller Method}
\label{sec:bm}
The Box-Muller method~\cite{box-muller} is a computational method
that generates samples of the standard normal distribution from uniformly distributed random values.
It operates as follows:

\begin{enumerate}
  \item Choose independent uniform random values $x_1$ and $x_2$ from $(0, 1)$
  {using $\RandomDouble$}.
  \item Set $r^2\leftarrow -2 \ln x_1$ and $\Theta \leftarrow 2 \pi x_2$.
  \item Return $s_1 \gets r \cos(\Theta)$.
\end{enumerate}
This procedure can be used to generate two independent normal samples:
$s_1$, as above, and $s_2 \gets r \sin (\Theta)$.

\begin{table*}[t]
\normalsize
  \centering
  \caption{The range of attack rate and accuracy across $\epsilon \in [1,20]$.
  Attack rate refers to the rate at which the attacker can make a gues.
  Attack accuracy measures how many of these guesses are correct.
  The two implementations of polar method and Box-Muller, $\gaussnumpy$ and $\gausspytorch$, 
  respectively, have different attack rates, though the accuracy of those
  guesses is always above 89\%. The attack rate and accuracy for $\gaussgo$ always 
  stay above $76\%$ and $99.9\%$, respectively.}
  \begin{tabular}{c|cc|cc|cc}
  \hline
  \multirow{2}{*}{Sensitivity} & \multicolumn{2}{c|}{$\gaussnumpy$}          & \multicolumn{2}{c|}{$\gausspytorch$}      & \multicolumn{2}{c}{$\gaussgo$}          \\ \cline{2-7} 
                                & attack rate         & attack accuracy      & attack rate          & {attack accuracy}  & attack rate          & attack accuracy      \\ \hline
 1                            & $[1.7\%, 78.2\%]$ & $[92.4\%, 99.9\%]$ & $[4.3\%, 92.8\%]$  & $[97.7\%, 99.9\%]$   & {[}76.4\%, 89.6\%{]} & {[}99.9\%, 100\%{)} \\
  10                           & $[1.9\%, 76.5\%]$ & $[89.6\%, 99.9\%]$ & $[10.9\%, 91.9\%]$ & {$[99.5\%, 99.9\%]$}    & {[76.3\%, 88.8\%{]}} & {[}99.9\%, 100\%{)} \\ \hline
  \end{tabular}
  \label{tab:recall}
\end{table*}

\subsection{DP Gradient Computation}
\label{app:gradcomp}
We now recall how $f$ and $\Delta_f$ are determined
for the DP-SGD mechanism. 
\begin{equation}
  f(B) = \frac{1}{S}\sum_{i=1}^{S} g(b_i)
\end{equation}
where $B= [b_1, b_2, \ldots, b_S]$ and $b_i$ is a record in $B$, $g$ calculates per-record gradient with per-record clipping with respect to clipping norm $L$.
Note that since the model has $d=26,010$ parameters and each batch is used to update them all,
$f(B)\in\mathbb{R}^d$ contains a gradient for each parameter of the model.

\paragraph{Sensitivity Analysis}
The sensitivity of DP-SGD is the maximum $L_2$ distance between any pair of
$f(B)$ and $f(B')$. Clipping norm $L$ dictates the largest 
$L_2$ norm of any record gradient.
Thus, for any $f(B)$ and $f(B')$, we can have:
\begin{equation*}
  \begin{aligned}
    \|f(B')-f(B)\|_2 & \le \frac{1}{S}\|g(b_c)-g(b_r)\|_2 \\
      & \le \frac{1}{S}(\|g(b_c)\|_2+\|g(b_r)\|_2)  \le \frac{2L}{S}
  \end{aligned}
\end{equation*}
where $b_c$ refers to the canary record and $b_r$ refers to the record replaced by $b_c$.
Since $L=1$ and {$S=64$}, the sensitivity of $f$ is $1/32$.

DP-SGD then proceeds by computing
\[ y = f(B) + \frac{1}{S}Z\]
where $Z = [Z_1,\ldots,Z_d] \stackrel{i.i.d.}{\sim} \mathcal{N}(0, \sigma^2L^2)$,
$y\in\mathbb{R}^d$,
and $\gausspytorch$  is used to draw noise from $\mathcal{N}$.
Note that even if some record is radically different from others in the batch (e.g., like
the canary used in $\difflabel$)
its gradient is clipped at $L$.

\subsection{DP-SGD Discretization}
\label{app:discmitig}
In Section~\ref{sec:deffp}, we  implement DP-SGD with discrete Gaussian by discretizing 
the gradient $g\in\mathbb{R}^d$ as described in~\cite{discrete-fl}
who use discrete Gaussian for training models in Federated Learning setting.
The method proceeds as follows.
\begin{enumerate}
  \item Clip $g$ w.r.t.~clipping norm $L$, $g' = g\times \min\{1, L/\left\lVert g\right\rVert_2\}$.
  \item Scale $g'$ with discretization 
  parameter $1/\gamma$, so the discretization can be done on a finer 
  grid $\gamma\mathbb{Z}^d$. Then randomly round each coordinate to the 
  nearest integer, $z=\mathsf{Round}_{\gamma}(g')$ so that
  $z\in\mathbb{Z}^d$.
  \item Let $G\in\mathbb{Z}^d$ consist of $d$ independent samples from the 
  discrete Gaussian $\mathcal{N}_{\mathbb{Z}}(0, \sigma^2/\gamma^2)$.
  \item Add discrete noise to $z$ and undo discretization, $z'=(z+G)\times \gamma$.
\end{enumerate}
where $\mathsf{Round}_{\gamma}$ is the conditional randomized rounding function with discretization
parameter $\gamma$. We refer the reader to~\cite{discrete-fl} 
for details about $\mathsf{Round}_{\gamma}$.

\subsection{Ablation study}
\label{app:recall}

Here we investigate the rate at which the attacker 
can make a guess (attack rate) and how many of these guesses are correct (attack accuracy).
To this end, we simulate settings where $q = 0$ and $q' = c$ and test two values of 
$c$: 1 and~10, meaning the sensitivity $\Delta$ of the underlying function is $1$ and $10$, respectively.
We take a range of values of~$\epsilon$ from~1 to~20 while keeping a fixed $\delta = 10^{-5}$.
Table~\ref{tab:recall}~shows that
implementations $\gaussnumpy$ and $\gausspytorch$ have different attack rates; 
however, their attack accuracy is always at least 89\%.
For $\gaussgo$, we observe that the attack rate is not influenced significantly by $\epsilon$ (at least 76\%), 
and the attack accuracy is always above $99.99\%$.

\subsection{Timing Attacks vs.~Probability-based Attacks}
\label{app:probattack}

\newcommand{\Geo}{\mathsf{Geometric}}
\newcommand{\Lap}{\mathsf{Laplace}}

In this section, we compare the effectiveness of timing attacks in Section~\ref{sec:e2e}
with a baseline based on the probability distribution (Laplace) of the underlying DP mechanism.

\subsubsection{Probability-based attack}
This baseline attack exploits the knowledge of the probability distribution from which DP noise
is sampled.
For example, for Laplace distribution with parameter $\sigma$ in~\cite{discrete}, the probability of drawing noise $s$
is $\frac{e^{1/\sigma}-1}{e^{1/\sigma} +1}e^{-|s|/\sigma}$.
The probability attack exploits this knowledge by choosing as its guess
a noise with the highest probability.

\subsubsection{Experimental setting}
{For attacks on all implementations in this section, we use the same settings as in~Section~\ref{sec:e2e}.} 
That is, we set the sensitivity $\Delta=5000$, and $\epsilon\in[1,10]$ and use
the private sum computation on two neighboring datasets $D$ and $D'$, with $\sum_{D} x=0$ and $\sum_{D'} x=5000$.
We ran 1 million trials to measure the success rates 
of timing and probability attacks for each $\epsilon\in[1,10]$.
For each trial, $D$ and $D'$ are chosen at random.

\subsubsection{Attacks on Laplace~\RNum{1}~\cite{discrete}}

\begin{figure}[t]
\centering
\includegraphics[width=8.7cm]{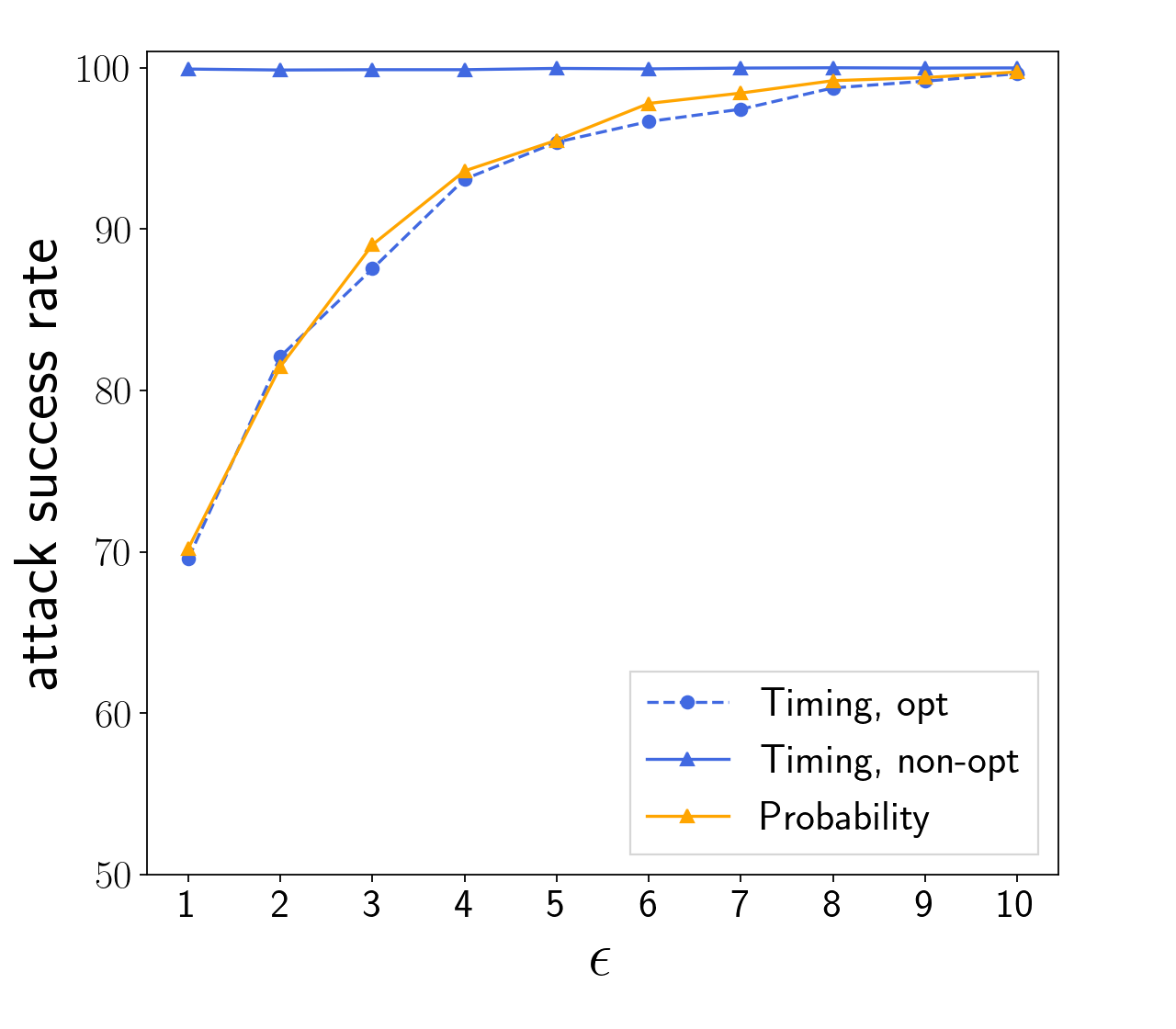}
\caption{
Attack success rates for timing and probability attacks,
on Laplace~\RNum{1}
where Geometric distribution sampling implementation is
optimized (opt) and non-optimized (non-opt).
$\epsilon\in[1,10]$, $\Delta=5000$.
The success rates for the probability attack are the same for opt and non-opt implementations.
}
\label{fig:eps_time_succ_impl1}
\end{figure}

We use mechanism Laplace~\RNum{1} based on
two versions of sampling from Geometric distribution in~\cite{discrete-imp}:
\begin{itemize}
    \item \texttt{sample\_geometric\_exp\_slow}: 
    non-optimized version. The algorithm
    samples noise from $\Geo(1-e^{-1/\sigma})$ with the parameter $\sigma$. 
    \item \texttt{sample\_geometric\_exp\_fast}:
    optimized version that aims to reduce the total number of operations for sampling~\cite[Proposition~34]{discrete}.
    It
    samples noise from $\Geo(1-e^{-1})$ and then
      scales it according to parameter $\sigma$. 
      \end{itemize}
We note that both of these implementations are also verified in Lean prover~\cite{lean-imp}.

The attack results against optimized and non-optimized versions appear in 
Figure~\ref{fig:eps_time_succ_impl1}.
We observe that the success rates for the timing attack against the optimized version is the same as the
probability attack. On the other hand, timing attack is much more effective against the
non-optimized version for $\epsilon \le 7$, with the success rate higher than 99\%.
The side-channel is more evident in the non-optimized version  since
the time it takes to draw the noise is linear in its magnitude. In the non-optimized version, the noise values are high since they are drawn from $\Geo(1-e^{-1/\sigma})$. The optimized version does not exhibit this dependence on the scale of the noise, since rescaling happens after the initial noise is drawn from $\Geo(1-e^{-1})$.

\subsubsection{Attacks on Laplace~\RNum{2}~\cite{google_gauss}}

\begin{figure}[t]
\centering
\includegraphics[width=8.7cm]{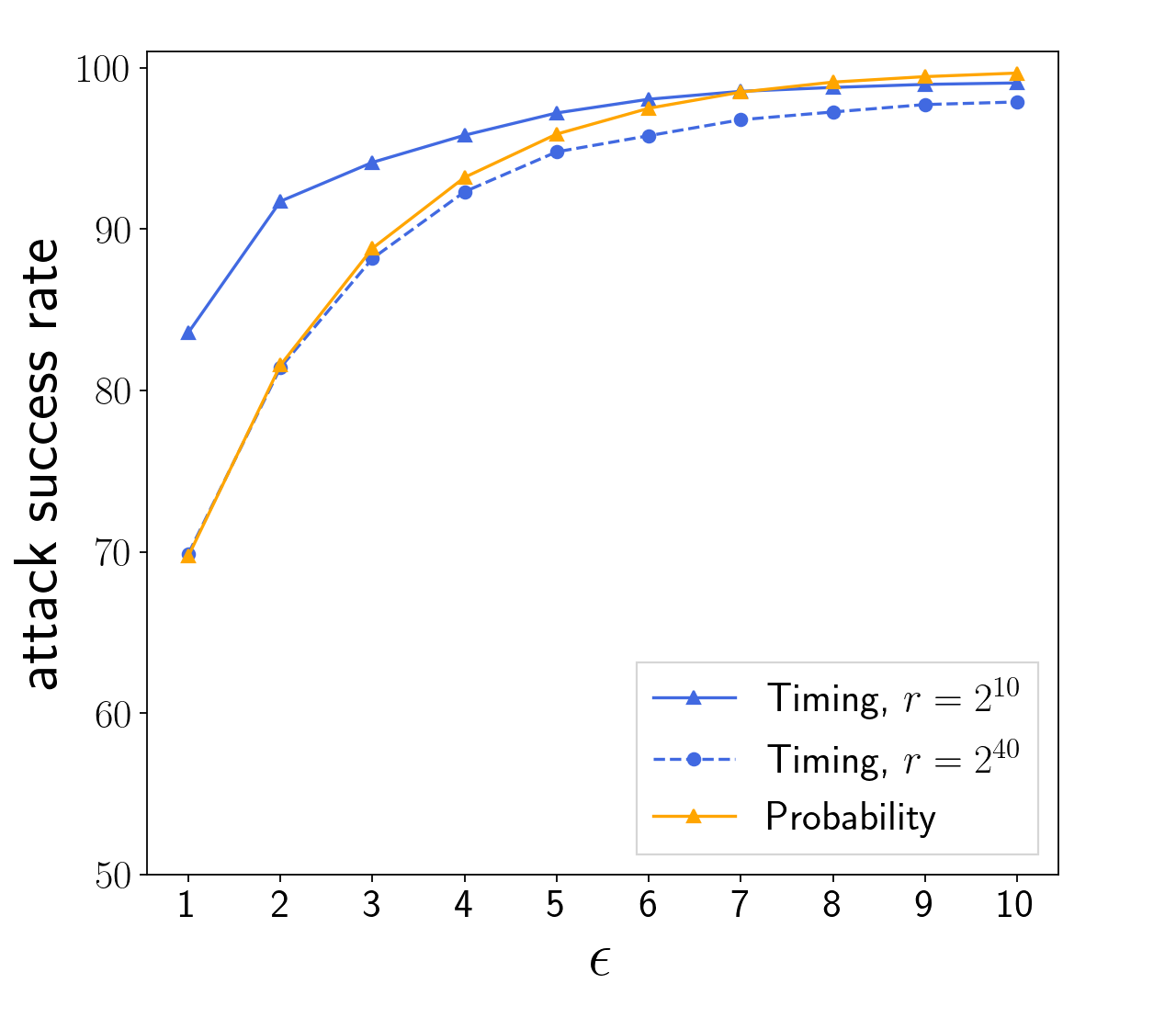}
\caption{
Attack success rates for timing and probability attacks on Laplace~\RNum{2},
with $\epsilon\in[1,10]$, $\Delta=5000$, and $r\in\{2^{10},2^{40}\}$.
The success rates for the probability attack are the same for 
$r=2^{10}$ and $r=2^{40}$.
}
\label{fig:eps_time_succ_impl2}
\end{figure}

We now test on the implementation Laplace~\RNum{2}. This implementation approximates Laplace mechanism
with  a resolution parameter $r=2^k$ that controls the expected relative error
and higher resolution leads to a smaller error~\cite[Proposition~2]{google_gauss}.

\begin{figure}[h]
\centering
\includegraphics[width=8.7cm]{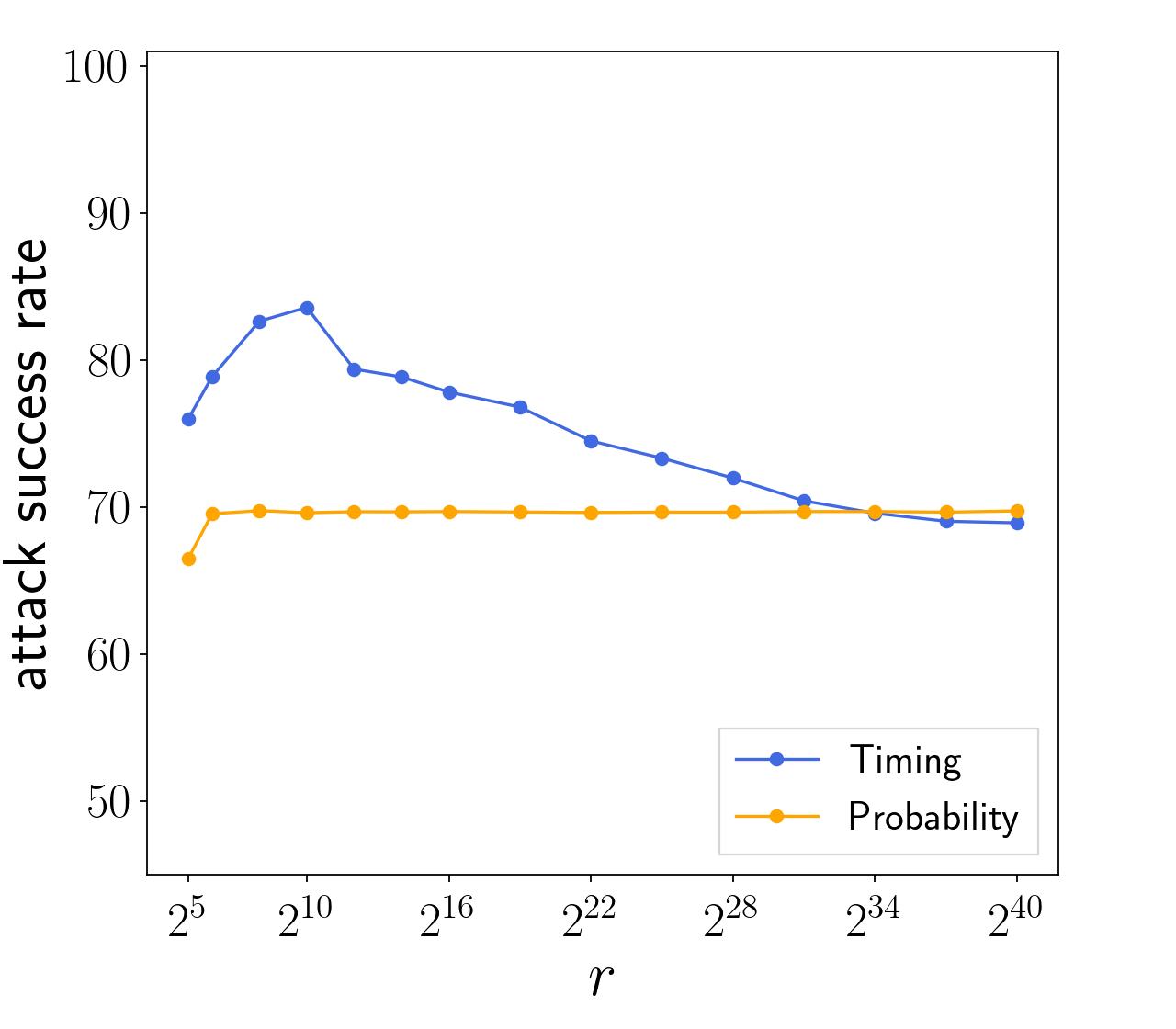}
\caption{
Attack success rates for timing and probability attacks on Laplace~\RNum{2},
with varying values of resolution $r$,
$\epsilon=1$ and $\Delta=5000$.
}
\label{fig:rsl_time_succ}
\end{figure}

We observe that for the default setting of $r=2^{40}$ the timing attack slightly underperforms the baseline probability attack
but outperforms it for $r=2^{10}$ (Figure~\ref{fig:eps_time_succ_impl2}).
To explore the influence of the resolution $r$ on 
the effectiveness of the timing attack,
we test on $r$ ranging from $2^5$ to $2^{40}$, 
with $\epsilon=1$ and $\Delta=5000$.
The attack results appear in Figure~\ref{fig:rsl_time_succ}.
Similar to the non-optimized version of  Laplace~\RNum{1}, albeit with different implementation (via binary search),
the timing of Laplace~\RNum{2} depends on the scale of the underlying Geometric distribution
which in this case depends on the resolution parameter~$r$.

\subsubsection{Summary}
In summary, 
the effectiveness of the timing attack depends on the implementation details,
including parameter settings and levels of optimization.
Depending on the implementation,
the performance of the timing attack is better
than or similar to that of the baseline probability attack.

\end{document}